\documentclass[conference]{IEEEtran}
\IEEEoverridecommandlockouts
\usepackage{cite}
\usepackage{amsmath,amssymb,amsfonts}
\usepackage{algorithmic}
\usepackage{graphicx}
\usepackage{textcomp}
\usepackage{xcolor}
\usepackage{url}
\usepackage{xspace}
\usepackage{booktabs}
\usepackage{multirow}
\usepackage[normalem]{ulem}
\useunder{\uline}{\ul}{}
\usepackage{color}
\usepackage{color, xspace}

\def\BibTeX{{\rm B\kern-.05em{\sc i\kern-.025em b}\kern-.08em
    T\kern-.1667em\lower.7ex\hbox{E}\kern-.125emX}}
\begin{document}

\title{SSDRec: Self-Augmented Sequence Denoising for Sequential Recommendation}

\author{{Chi Zhang\textsuperscript{1}, Qilong Han\textsuperscript{*,1}\thanks{*Corresponding authors}, Rui Chen\textsuperscript{1}, Xiangyu Zhao\textsuperscript{*,2}, Peng Tang\textsuperscript{3}, Hongtao Song\textsuperscript{1}}\\
\textsuperscript{1}\textit{College of Computer Science and Technology, Harbin Engineering University} \\
\textsuperscript{2}\textit{School of Data Science, City University of Hong Kong} \\
\textsuperscript{3}\textit{School of Cyber Science and Technology, Shandong University} \\
\{zhangchi20,hanqilong,ruichen,songhongtao\}@hrbeu.edu.cn, xianzhao@cityu.edu.hk, tangpeng@sdu.edu.cn}

\maketitle

\newcommand{\model}{SSDRec\xspace}
\newcommand{\problem}{OUPs\xspace}

\newcommand{\framework}{\textbf{\underline{S}}elf-augmented \textbf{\underline{S}}equence \textbf{\underline{D}}enoising for sequential \textbf{\underline{Rec}}ommendation (\model)\xspace}

\begin{abstract}
Traditional sequential recommendation methods assume that users' sequence data is clean enough to learn accurate sequence representations to reflect user preferences. In practice, users' sequences inevitably contain noise (e.g., accidental interactions), leading to incorrect reflections of user preferences. Consequently, some pioneer studies have explored modeling sequentiality and correlations in sequences to implicitly or explicitly reduce noise's influence. However, relying on only available intra-sequence information (i.e., sequentiality and correlations in a sequence) is insufficient and may result in over-denoising and under-denoising problems (\problem), especially for short sequences. To improve reliability, we propose to augment sequences by inserting items before denoising. However, due to the data sparsity issue and computational costs, it is challenging to select proper items from the entire item universe to insert into proper positions in a target sequence. Motivated by the above observation, we propose a novel framework--Self-augmented Sequence Denoising for sequential Recommendation (\model) with a three-stage learning paradigm to solve the above challenges. In the first stage, we empower \model by a global relation encoder to learn multi-faceted inter-sequence relations in a data-driven manner. These relations serve as prior knowledge to guide subsequent stages. In the second stage, we devise a self-augmentation module to augment sequences to alleviate \problem. Finally, we employ a hierarchical denoising module in the third stage to reduce the risk of false augmentations and pinpoint all noise in raw sequences. Extensive experiments on five real-world datasets demonstrate the superiority of \model over state-of-the-art denoising methods and its flexible applications to mainstream sequential recommendation models. The source code is available online at \url{https://github.com/zc-97/SSDRec}.
\end{abstract}

\section{Introduction}
\label{sec:introduction}
Recommender systems have shown their power in boosting business revenue and improving user experience in many real-world applications, such as news services~\cite{ZZST19}, social media~\cite{RLL17,ZQD19}, and e-commerce platforms~\cite{ZZS18,ZXY20}. Traditional collaborative filtering techniques~\cite{RFZ09, HPK16, YHC18, WHW19, HDW20, MZX21} mainly rely on pairwise user-item interactions and ignore the natural temporal sequentiality in users' interaction sequences. This limitation hinders their ability to comprehensively capture users' time-evolving preferences in practical recommendation scenarios, which motivates the line of research on \textit{sequential recommendation}.

Sequential recommendation aims to recommend the most likely next item to a target user based on her historical interaction sequence. Effectively representing a user's interaction sequence to reflect her time-evolving preferences lies at the core of sequential recommendation. Therefore, recent sequential recommendation studies~\cite{HKB15,DLZ17,LRC17,TW18,KM18,SLW19,WTZ19} leverage various deep models (e.g., recurrent neural networks~\cite{HKB15,DLZ17}, convolutional neural networks~\cite{TW18}, Transformer~\cite{KM18,LRC17,SLW19}, and graph neural networks~\cite{WTZ19}) to capture the sequentiality in sequences, which is then used to learn sequence representations and make recommendations. However, in practice users' interaction sequences inevitably contain \textit{inherent noise} (e.g., accidental interactions~\cite{TLF19, WWQ22,ZCZH23}), which blurs users' real preferences. Learning sequence representations from such noisy sequences can ultimately harm the performance of sequential recommenders, calling for the need of \textit{sequence denoising}. However, despite its potential benefits, the lack of supervised labels to indicate noise makes the problem of sequence denoising technically challenging.

There have been some recent studies that explore various denoising methods to refine raw sequence data and improve data utility (e.g., explicitly remove noise~\cite{WFH21,YSS21,QWL21,TWL21,SWS21,ZDZ22,ZCZH23} or implicitly attenuate the influence of noise in sequences~\cite{ZYZ22}), thus enhancing the quality of learned representations. The key assumption of the existing methods is that a noiseless sequence typically exhibits \textit{smooth sequentiality} (i.e., each item exhibits sequential relations with its context in a sequence) and \textit{high correlation} (i.e., each item is relevant with the sequence's next interaction in terms of item-wise similarity~\cite{ZDZ22}), while noisy sequences do not. Therefore, it is possible to leverage the information learned from a single sequence to identify noise that disrupts its sequentiality and correlations. Despite their effectiveness of addressing the sequence denoising problem, it is crucial to recognize that the information available in a single sequence is limited, especially in short sequences in many applications. Since the observed interactions in a single sequence are usually limited and there is a lack of labels to indicate noise, the learned sequential information and correlations may be insufficient to accurately denoise and even disrupt advantageous information. Overlooking this limitation may result in unreliable denoising results, leading to \textit{over-denoising} and \textit{under-denoising} problems (\problem), which can inevitably make sequential recommenders sub-optimal.

\begin{figure}[t]
\centering
\setlength{\abovecaptionskip}{1mm}
  \includegraphics[width=\linewidth]{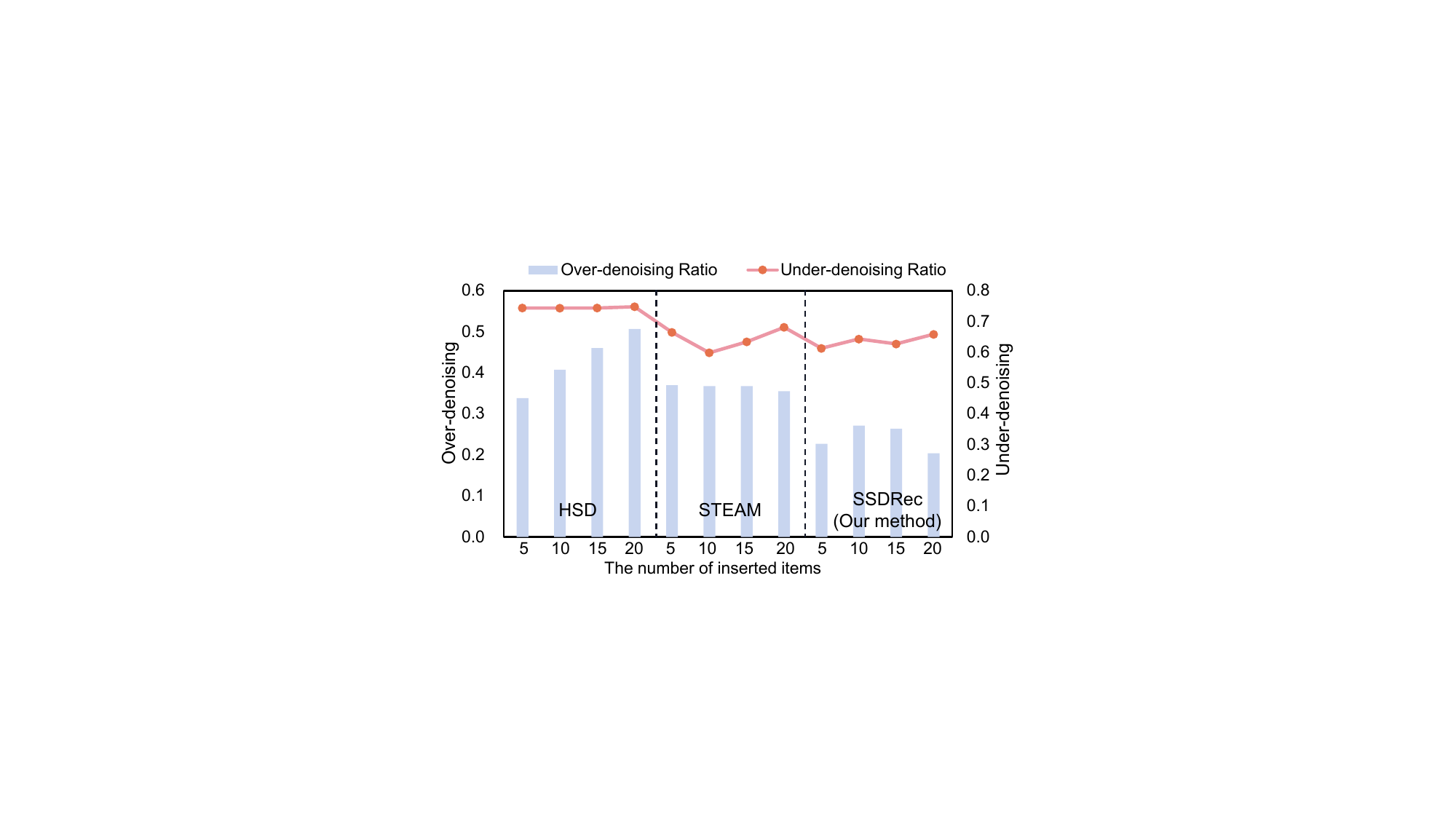}
\caption{The \problem of different sequence denoising methods on ML-100K.}
\label{fig: motivation}
\end{figure}

As illustrated in Figure~\ref{fig: motivation}, we randomly insert unobserved interactions as noise into raw short sequences (filtering out low-rating items with ratings less than 3) to verify the existence of \problem. We consequently calculate how many inserted items will be kept and how many raw items will be dropped to represent the under-denoising ratio and over-denoising ratio, respectively. It can be observed that HSD~\cite{ZDZ22} and STEAM~\cite{LWC23} (two representative explicit denoising methods learning intra-sequence information) suffer from \problem. Therefore, to alleviate \problem, we design a self-augmentation method to enrich short sequences before denoising and reduce the risk of introducing false augmentations. However, since interactions in a single sequence are limited compared to the entire item universe, it is difficult to match and rank proper items without additional features to guide sequence augmentation. Moreover, the complexity of inserting items could significantly increase with the problem scale (i.e., sequence lengths), and it follows that it is impractical to insert items to each position in a sequence. Thus, achieving a reasonable trade-off between performance and efficiency renders two unique technical challenges: (1) \textit{Which items to select}? (2) \textit{Which positions in a target sequence to choose}?


\begin{figure*}[ht]
  \centering
  \setlength{\abovecaptionskip}{1mm}
    \includegraphics[width=\linewidth]{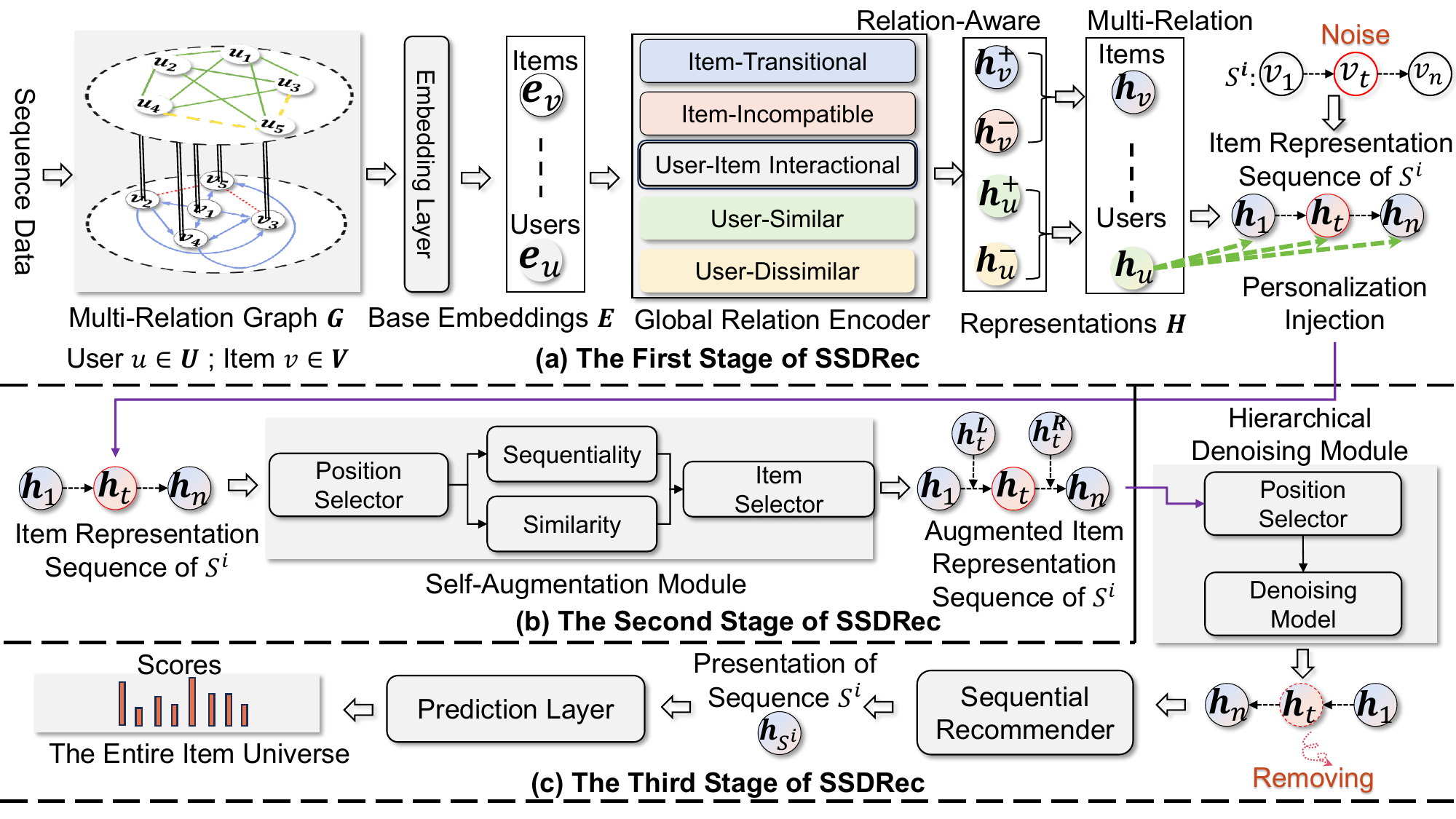}
  \caption{The architecture of the proposed \model framework. \textcolor{black}{ (a) The first stage of \model takes the multi-relation graph $\mathcal{G}$ as input and outputs the item representation sequence for subsequent denoising stages; (b) The second stage of \model selects items and positions to augment sequences; (c) The third stage of \model adopts a hierarchical denoising module to ensure augmentation reliability and generate noiseless sequences for making recommendations. 
  }
  }
 \label{fig: model}
\end{figure*}

\noindent\textbf{Contributions.} In view of the above challenges, we propose a novel framework--\framework with a unique three-stage denoising paradigm. Specifically, in the first stage, we devise a global relation encoder to learn multi-faceted relations of users and items in a data-driven manner. As a result, the learned relations serve as inter-sequence prior knowledge to guide the following stages. In the second stage, we propose a self-augmentation module to select proper items and positions to generate augmented sequences. Technically, we design a position selector that detects sequentiality and similarity inconsistencies to select a position in the target sequence. After that, we put forward an item selector to match and rank items for insertion if the sequence is short. The self-augmentation module can better the subsequent denoising process and avoid introducing too much complexity compared to inserting as many items as possible. Finally, we adopt a hierarchical denoising module to identify all potential noise in the third stage. The hierarchical denoising module first refines the augmented sequence by filtering out false augmentations introduced in the second stage and then gradually pinpoints all potential noise in the target sequence, thus enhancing denoising reliability. We further provide a case study in Section~\ref{subsec:case_study} to show how the \model framework alleviates \problem and affects recommendations.

Our main technical contributions are summarized as follows.
\begin{itemize}[\IEEElabelindent=0pt]
    \item To the best of our knowledge, this is the first paper that proposes to ameliorate sequence denoising from a new perspective of tackling \problem by performing explicit sequence augmentation before denoising. We identify two critical technical challenges in augmentation, which have not been previously explored in the literature. 
    \item We propose a novel sequence denoising framework \model, consisting of three stages to alleviate \problem and tackle augmentation challenges. We empower \model without requiring any labels for noise or additional features, which satisfies the settings of most existing sequential recommendation methods. Thus, the proposed solution can seamlessly serve as a plug-in to reliably generate noiseless sub-sequences for most existing sequential recommenders.
    \item We perform extensive experiments on five real-world public recommendation datasets and demonstrate the superiority of the proposed \model framework over state-of-the-art denoising methods and its flexible and effective applications to mainstream sequential recommendation models.
\end{itemize}

\section{Preliminaries}
\label{sec:preliminaries}
\begin{table}[t]
\centering
\caption{Descriptions of Key Notations}
\label{tab: notations}
\resizebox{\columnwidth}{!}{%
\begin{tabular}{lll}
\toprule
Notations & Description &  \\ \midrule
$\mathcal{G}$ & \multicolumn{2}{l}{The constructed multi-relation graph.} \\
$u,v$ & \multicolumn{2}{l}{User and item.} \\
$d$ & \multicolumn{2}{l}{Embeddings' dimension size.} \\
$S,s_t,n_i$ & \multicolumn{2}{l}{The target sequence, the $t$-th item in $S$, and $S$'s length.} \\
$\boldsymbol{e}_u,\boldsymbol{e}_v\in\mathbb{R}^d$ & \multicolumn{2}{l}{Embeddings of user $u$ and item $v$.} \\
$\boldsymbol{h}_{v}^{+},\boldsymbol{h}_{v}^{-},\boldsymbol{h}_{u}^{+},\boldsymbol{h}_{u}^{-}\in\mathbb{R}^d$ & \multicolumn{2}{l}{Relation-aware representations of different relations.} \\
$\boldsymbol{h}_{u},\boldsymbol{h}_v\in\mathbb{R}^{d}$ & \multicolumn{2}{l}{Multi-relation representations of user $u$ and item $v$.} \\
$\boldsymbol{\hat{r}}_{s_t}\in\{0,1\}$ & \multicolumn{2}{l}{The hard-coding inconsistency score of $s_t$ in $S$.} \\
$\boldsymbol{h}^{L}_{t},\boldsymbol{h}^{R}_{t}\in\mathbb{R}^d$ & The inserted items. &  \\\bottomrule
\end{tabular}%
}
\end{table}

In this section, we introduce key definitions in this work and summarize frequently used notations in  Table~\ref{tab: notations}.

\noindent\textbf{User-Item Interaction Data.}
Let $\mathcal{U}=\{u_1,u_2,\cdots,u_{\vert \mathcal{U} \vert}\}$ and $\mathcal{V}=\{v_1,v_2,\cdots,v_{\vert \mathcal{V} \vert}\}$ be the set of users and items, respectively. We define the user-item interaction data as an interaction matrix $\boldsymbol{A} \in\mathbb{R}^{\vert \mathcal{U} \vert \times \vert \mathcal{V} \vert}$. Each element in $\boldsymbol{A}$ indicates how many times user $u$ has interacted with item $v$.

\noindent\textbf{Raw Sequence Data.} Following previous studies~\cite{HKB15,DLZ17,LRC17,TW18,KM18,SLW19,WTZ19}, we use $S^i=[s_{1}^i,\cdots,s_{t}^i,\cdots,s_{n_i}^i]$ and $s_{n_i+1}^i$ to denote user $u_i$'s historical temporal interaction sequence and her next interaction, respectively. $s_{t}^i \in \mathcal{V}$ is the $t$-th item interacted by user $u_i$, and $n_i$ represents the sequence's length. 

\noindent\textbf{Sequential Recommendation.} For each user $u_i$, sequential recommendation takes $u_i$'s raw sequence data $S^i$ as input and outputs the top-$k$ items from $\mathcal{V}$ that are most likely to be interacted at the next time step of $S^i$, i.e., $n_i+1$.

\noindent\textbf{Sequence Denoising.} Given a sequence $S^i$ of user $u_i$, explicit sequence denoising methods~\cite{YSS21, ZDZ22} focus on attenuating the negative influence of noise (i.e., removing identified noise) in $S_i$, thus generating a noiseless sub-sequence $S^i_-=[s_{1-}^i,\cdots,s_{t-}^i,\cdots,s_{n_{i-}}^i]$, where $s_{t-}^i\in S^i$ and $n_{i-}\leq n_i$.

\noindent\textbf{Problem Statement.} We formally define our task as follows: 
\textbf{Input}: A triplet set $\mathcal{S}=\{(u_i, S^i,s^i_{n_i+1})\vert u_i\in\mathcal{U}\}$, which contains triplets consisting of user $u_i$, $u_i$'s raw interaction sequence $S^i$, and $u_i$'s next interaction $s^i_{n_i+1}$.

\noindent\textbf{Output}: The predictive function $\mathcal{F}(u_i,v_i\vert \mathcal{S}, \Theta)$, which estimates the likelihood of each user $u_i$ adopting an item $v_i\in\mathcal{V}$ as her next interaction, where $\Theta$ is the set of model parameters. The optimized parameters (i.e., $\Theta^*$) is learned via an end-to-end learning process that generates noiseless sequences (i.e., $S^i_-$) and maximizes the likelihood of recommending item $s^i_{n_i+1}$ for each user $u_i$, denoted by $\mathcal{P}(\mathcal{F}(u_i,s^i_{n_i+1}\vert \mathcal{S}, \Theta^*))$.

\section{Methodology}
\label{sec:methodology}

As illustrated in Figure~\ref{fig: model}, \model is powered by a three-stage learning paradigm. The first stage (Figure~\ref{fig: model}(a)) is equipped with a global relation encoder to capture the relations among users and items. The relations serve as prior knowledge to guide the following stages. We consequently propose a self-augmentation module in the second stage (Figure~\ref{fig: model}(b)) to select items and positions to augment sequences. Finally, we adopt a hierarchical denoising module (Figure~\ref{fig: model}(c)) to avoid false augmentations and perform sequence denoising.

\begin{figure}[t]
\centering
\setlength{\abovecaptionskip}{1mm}
  \includegraphics[width=\linewidth]{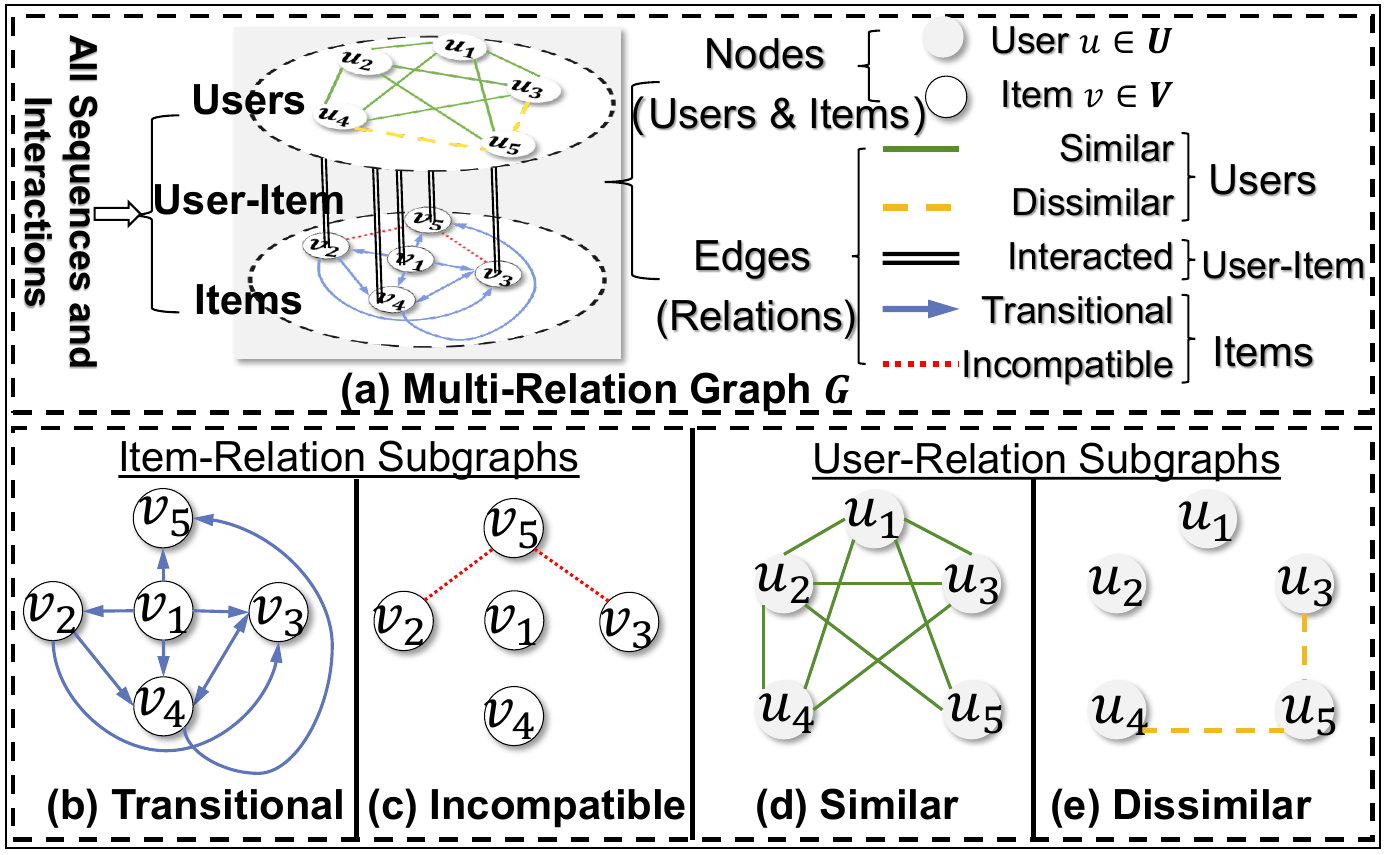}
\caption{A toy example of the construction of a multi-relation graph.}
\label{fig: graphs}
\end{figure}

\subsection{Multi-Relation Graph Construction}
In this section, we construct a multi-relation graph $\mathcal{G}$ in a data-driven manner, serving as prior knowledge to guide the following stages, e.g., select items and positions for sequence augmentation. As shown in Figure~\ref{fig: graphs}(a), the graph $\mathcal{G}=(\mathcal{N},\mathcal{E})$ comprises two types of nodes: users and items, and five types of relations: \textit{similar} and \textit{dissimilar} users, \textit{interacted} user-item, and \textit{transitional} and \textit{incompatible} items. The node set $\mathcal{N}=\mathcal{U}\cup\mathcal{V}$ contains all users and items. The edge set $\mathcal{E}=(\mathcal{E}^{+}_{vv},\mathcal{E}^{-}_{vv},\mathcal{E}_{uv},\mathcal{E}^{+}_{uu},\mathcal{E}^{-}_{uu})$ represents different relations among users (u) and items (v). In particular, we construct user-item interactional relations (i.e., $\mathcal{E}_{uv}$) by user-item interaction data $\boldsymbol{A}$ to connect user and item relations for better relation learning. There is an edge $\varepsilon_{uv}$ between user $u$ and item $v$ iff $\boldsymbol{A}_{uv}\neq 0$. To indicate interactional strength, we formulate the weight of $\varepsilon_{uv}$ as $w_{uv}=\boldsymbol{A}_{uv}$.

\subsubsection{Item-Relation Sub-Graphs}
Following previous studies~\cite{WZM20,HZC22}, we consider two types of relations: \textit{transitional} and \textit{incompatible} (i.e., $\mathcal{E}^{+}_{vv},\mathcal{E}^{-}_{vv}\in\mathcal{G}$), to capture global positive and negative dependencies among items, respectively. 

{\textbf{Transitional Relations.}} Transitional relations are unidirectional, which reflect the sequential co-occurrence of two items (e.g., a mobile phone and phone cases in e-commerce) in sequences, representing items' \textit{positive} dependencies. As illustrated in Figure ~\ref{fig: graphs}(b), there is an edge $\varepsilon^{+}_{ij}$ from items $v_i$ to $v_j$ iff $\text{Cnt}(v_i\rightarrow v_j)>0$, where $\text{Cnt}(\cdot)$ is a counter. To indicate the transitional strength between two items, we calculate the weight $w^{+}_{ij}$ of $\varepsilon^{+}_{ij}$ by jointly considering the positions and co-occurrence times of the two items in each user $u_i$'s sequence $S^i$. Specifically, we use the formula $\sum_{u_i}{\frac{n_i-\text{Dis}(v_i,v_j)}{n_i}}$ to carry out over all users that interact with items $v_i$ and $v_j$, where $\text{Dis}(\cdot,\cdot)$ is a calculator recording the positional distance of two items in $S^i$, and $n_i$ is the length of $S^i$. 

Accordingly, we can control the inserting probability of two transitional items. Having the former in a specific position in a sequence will increase the probability of inserting the latter after the position while decreasing the probability of inserting the latter before the position into the sequence.

{\textbf{Incompatible Relations.}} 
In contrast with transitional relations, incompatible relations are undirected, indicating \textit{negative} dependencies of two items (e.g., different generations of products that are not compatible with each other). Accordingly, we can reduce false augmentations for two incompatible items because having one of them in a sequence will lower the probability of inserting the other in the sequence. As shown in Figure~\ref{fig: graphs}(c), following the definition in MGIR~\cite{HZC22}, we construct incompatible relations between popular items (avoiding unreliable relations between long-tail items) by considering whether they ever co-exist in any sequences while having the same context, i.e., having transitional relations with at least one common item. Formally, an edge $\varepsilon^{-}_{ij}$ exists between two popular items $v_i$ and $v_j$ iff the items have a nonempty common item set $\mathcal{V}_{k}\neq \emptyset$, where $\mathcal{V}_k = \{v_k \vert (w^{+}_{ik}+w^{+}_{ki})\cdot (w^{+}_{jk}+w^{+}_{kj})\neq0\}\subsetneqq\mathcal{V}$, and they have no transitional relations, i.e., $\varepsilon^{+}_{ij}\cup\varepsilon^{+}_{ji}=\emptyset$. To indicate the incompatible strength between the two items, we calculate the weight $w^{-}_{ij}$ of $\varepsilon^{-}_{ij}$ via $\sum_{v_k}{(w^{+}_{ik}+w^{+}_{ki}+w^{+}_{jk}+w^{+}_{kj})}$, where $v_k\in\mathcal{V}_k$.

\subsubsection{User-Relation Sub-Graphs}
\label{subsubsec:user graph}
In practice, most long-tail items have never co-occurred, and the definition of incompatible relations does not include these items. However, there is no doubt that users' preferences are diverse. Some users prefer to interact with niche items. Selecting items from the entire item universe (typically containing many long-tail items) for the target sequence is difficult. To tackle this limitation, we propose constructing user relations to control item relation reliability under personalized implicit constraints. Specifically, we consider two types of user relations to construct user-relation sub-graphs: similar ($\varepsilon^{+}_{uu}$) and dissimilar ($\varepsilon^{-}_{uu}$).

{\textbf{Similar Relations.}} Functionally, similar relations are undirected, indicating that two users with similar interactive patterns prefer to interact with identical items. 
As illustrated in Figure~\ref{fig: graphs}(d), we use users' co-interactions to learn the similar relations. Mathematically, an edge $\varepsilon^{+}_{ij}$ exists between user $u_i,u_j\in\mathcal{U}$ iff the users have a nonempty common item set $\mathcal{V}_{k}=\{v_k\vert w_{ik}\cdot w_{jk}\neq0\}\subset\mathcal{V}$. We calculate the weight $w^{+}_{ij}$ of $\varepsilon^{+}_{ij}$ by Jaccard similarity as $\frac{\sum_{v_k}({w_{ik}+w_{jk}})}{\sum({w_{i\cdot}+w_{j\cdot}})}$.

{\textbf{Dissimilar Relations.}} The dissimilar relations between two users mean that the users' interactive patterns are individual, i.e., they never co-interact with the same items. Motivated by the definition of incompatible relations between items, as illustrated in Figure~\ref{fig: graphs}(d), we consider two users are dissimilar if they never co-interact with same items but have at least an identically similar user. 
Mathematically, there exists an edge $\varepsilon^{-}_{ij}$ between two users $u_i,u_j\in\mathcal{U}$ iff they share a common similar user set $\mathcal{U}_k\neq\emptyset$, where $\mathcal{U}_k = \{u_k \vert w^{+}_{ik}\cdot w^{+}_{kj}\neq0\}\subsetneqq\mathcal{U}$, and they are not similar users, i.e., $\varepsilon^{+}_{ij}=\emptyset$. We calculate the weight $w^{-}_{ij}$ of $\varepsilon^{-}_{ij}$ via $\sum_{u_k}{(w^{+}_{ik}+w^{+}_{kj})}$, where $u_k\in\mathcal{U}_k$.



\subsection{Embedding Layer}
\label{subsec: embedding layer}
To train \model, we first learn items' and users' embeddings by mapping their IDs to dense embedding vectors $\boldsymbol{e}_v,\boldsymbol{e}_u \in \mathbb{R}^d$. Consequently, we build two embedding look-up tables for embedding initialization as follows
\begin{equation}
    \boldsymbol{e}_{v} = \boldsymbol{x}_{v}\boldsymbol{W}_{v} \: ; \:
    \boldsymbol{e}_{u} = \boldsymbol{x}_{u}\boldsymbol{W}_{u},
    \label{eq:embedding}
\end{equation}
where $\boldsymbol{W}_{v} \in \mathbb{R}^{\vert \mathcal{V} \vert \times d}$ and $\boldsymbol{W}_{u} \in \mathbb{R}^{\vert \mathcal{U} \vert \times d}$ are trainable parameter matrices. $\boldsymbol{x}_{v}$ and $\boldsymbol{x}_{u}$ are one-hot encoding vectors of the IDs of item $v$ and user $u$, respectively. 

\subsection{Global Relation Encoder}
\label{sebsec: relation encoder}
As illustrated in Figure~\ref{fig: model}(a), we devise a global relation encoder to individually encode different relations in $\mathcal{G}$ by different relation encoding layers. 
\subsubsection{Item-Transitional Relation Encoding Layer} Since the transitional relations of items are unidirectional, we have to learn different neighbors' (i.e., incoming and outgoing) sequential dependencies. Accordingly, we devise a message passing scheme~\cite{VCC18,HZC22,ZCZH23} powered by the attention mechanism to distinguish transitional items' dependencies. 
Technically, for each item $v$ in $\mathcal{G}$ exhibiting transitional relations, we first assign a $2$-dimensional attention vector $\boldsymbol{\alpha}=[\alpha_{i}, \alpha_{j}]$ to weigh the contributions of $v$'s incoming (i.e., $\mathcal{V}^{+}_{i}=\{v_i\vert \varepsilon^{+}_{v_iv}\neq\emptyset,v_i\in\mathcal{V}\}$) and outgoing (i.e., $\mathcal{V}^{+}_{j}=\{v_j\vert \varepsilon^{+}_{vv_j}\neq\emptyset,v_j\in\mathcal{V}\}$) neighbors as follows:
\begin{equation}
\begin{aligned}
    \boldsymbol{\alpha}\!=\!\! \rho(\sigma(\boldsymbol{e}_{v} \!\boldsymbol{W}^{+}_{v_iv}\sum_{v_i}w^{+}_{v_iv}\boldsymbol{e}^{\mathsf{T}}_{v_i}) \Vert \sigma(\boldsymbol{e}_{v} \!\boldsymbol{W}^{+}_{vv_j}\!\!\sum_{v_j}\!\! w^{+}_{vv_j}\!\!\boldsymbol{e}^{\mathsf{T}}_{v_j})),
\end{aligned}
\end{equation}
where $\rho$ is a score function (e.g., the scaling and Softmax operators in Transformer~\cite{VSP17}), $\sigma(\cdot)$ is an activation function (e.g., the ReLU function), $\Vert$ is the concatenation operation, and $\boldsymbol{W}^{+}_{v_iv},\boldsymbol{W}^{+}_{vv_j}\in\mathbb{R}^{d\times d}$ are trainable matrices denoted by $\Theta_{att}=\{\boldsymbol{W}^{+}_{v_iv},\boldsymbol{W}^{+}_{vv_j}\}$. 
After that, we aggregate information of $v$ and its neighbors to learn $v$'s transitional-aware representation $\boldsymbol{h}^{+}_{v}$ by a convolution operator, which can effectively and efficiently learn homogeneous relations. Specifically, we calculate $\boldsymbol{h}^{+}_{v}$ as follows
\begin{equation}
\boldsymbol{h}^{+}_{v}=f^{+}_{v}(\alpha_i\sum_{v_i} w^{+}_{v_iv}\boldsymbol{e}^{\mathsf{T}}_{v_i}+\alpha_j \sum_{v_j} w^{+}_{vv_j}\boldsymbol{e}^{\mathsf{T}}_{v_j})\Vert \boldsymbol{e}_{v}\vert \Theta^{+}_{v}),
\label{eq: item positive agg}
\end{equation}
where $f^{+}_{v}$ is a convolution operator with stride 1 and filter size $2 \times 1$, $\Theta^{+}_{v}$ is the set of trainable weight parameters of the aggregation layer, i.e., convolutional filter. 

\subsubsection{Item-Incompatible Relation Encoding Layer} In contrast with item-transitional relations, the incompatible relations between two items are undirected. Therefore, we have no need to distinguish $v$'s neighbors, but straightly leverage a similar convolution used in Eq.~(\ref{eq: item positive agg}) with different parameters ($\Theta^{-}_{v}$) to learn $v$'s incompatible-aware representation $\boldsymbol{h}^{-}_{v}$ via:
\begin{equation}
        \boldsymbol{h}^{-}_{v} =f^{-}_{v}(\sum_{v_i}w^{-}_{v_iv}\boldsymbol{e}_{v_i}\Vert \boldsymbol{e}_v,\Theta^{-}_{v}).
\label{eq: item incompatible agg}
\end{equation}

\subsubsection{User-Item Interactional Relation Encoding Layer} Due to the inherent heterogeneity of user-item relations (i.e., containing user and item nodes), we adopt a lightweight message passing function in DPT~\cite{ZCZH23} based on LightGCN\cite{HDW20} to connect users and items. Formally, for each item $v$ and user $u$, we learn their representations as
\begin{equation}
        \boldsymbol{h}^{u\rightarrow v}_{v} = \sum_{u_i}(w_{u_iv}\boldsymbol{e}_{u_i})\: ;\: \boldsymbol{h}^{v\rightarrow u}_{u} = \sum_{v_j}(w_{v_ju}\boldsymbol{e}_{v_j}),
\label{eq: user-item agg}
\end{equation}
where $\mathcal{U}_i=\{u_i \vert \varepsilon_{u_iv}\neq\emptyset,u_i\in\mathcal{U}\}$ is the set of users that have interacted with item $v$, and $\mathcal{V}_j=\{v_j \vert \varepsilon_{v_ju}\neq\emptyset,v_j\in\mathcal{V}\}$ is the set of items that are interacted by user $u$.
\subsubsection{User-Similar Relation Encoding Layer} To learn users' similar interaction patterns, for each user $u$, we learn $u$'s similar-aware representation $\boldsymbol{h}^{+}_{u}$ as follows
\begin{equation}
\begin{aligned}
    \boldsymbol{h}^{+}_{u}=f^{+}_{u}(\sum_{u_i}w^{+}_{u_iu}\boldsymbol{e}_{u_i}\Vert \boldsymbol{e}_u,\Theta^{u+}_{u}),
\end{aligned}
\label{eq: user similar agg}
\end{equation}
where $\mathcal{U}^{+}_{i}=\{u_i\vert \varepsilon^{+}_{u_iu}\neq\emptyset\}$ is $u$'s neighbor containing her similar users in $\mathcal{G}$, and $f^{+}_{u}$ is a similar convolution function used in Eq.~(\ref{eq: item positive agg}) and Eq.~(\ref{eq: item incompatible agg}) with different parameters $\Theta^{u+}_{u}$.
\subsubsection{User-Dissimilar Relation Encoding Layer} For each user $u$, we learn $u$'s dissimilar-aware representations $\boldsymbol{h}^{-}_{u}$ via
\begin{equation}
    \boldsymbol{h}^{-}_{u} =f^{-}_{u}(\sum_{u_i}w^{-}_{u_iu}\boldsymbol{e}_{u_i}\Vert \boldsymbol{e}_u,\Theta^{-}_{u}),
\label{eq: user dissimilar agg}
\end{equation}
where $\mathcal{U}^{-}_{i}=\{u_i\vert \varepsilon^{-}_{u_iu}\neq\emptyset\}$ is the set of $u$'s dissimilar users in $\mathcal{G}$, and $\Theta^{-}_{u}$ is the set of parameters of the convolution function used in Eq.~(\ref{eq: user similar agg}) with different parameters. While this layer is similar to the item-incompatible relation encoder, i.e., Eq.~(\ref{eq: item incompatible agg}), it can implicitly reduce some items' relevance from a user perspective to better global relation learning. For example, this is true for long-tail items, which have never co-existed in any sequences and do not exhibit incompatible relations.

With the learned representations, we merge relation-specific information to learn multi-relation representations as
\begin{equation}
    \boldsymbol{h}_{v} = f_{v}(\boldsymbol{h}^{+}_{v},\boldsymbol{h}^{-}_{v},\boldsymbol{h}^{u\rightarrow v}_{v}\vert\Theta_{v}) ; \boldsymbol{h}_{u} = f_{u}(\boldsymbol{h}^{+}_{u},\boldsymbol{h}^{-}_{u},\boldsymbol{h}^{v\rightarrow u}_{u}\vert\Theta_{u}).
\end{equation}
For instance, we adopt two feedforward layers with different trainable parameters (i.e., $\Theta_{v}$ and $\Theta_{u}$) to implement the fusion functions to generate multi-relation representations $\boldsymbol{h}_v$ and $\boldsymbol{h}_u$. 
\subsection{Self-Augmentation Module}
\label{subsee:self-augmentaion}
With the global relation encoder, we can inject sufficient inter-sequence information to select items and positions for sequence augmentation. To fully capture the complex relations among items and users, we first fuse users' and items' multi-relation representations to generate an informative item representation sequence $\boldsymbol{H}_S$ of the target sequence $S$. 
Specifically, for each item (i.e., $s_t=v\in\mathcal{V}$) in $\boldsymbol{H}_S$ interacted by user $u$, we generate its representation via $\boldsymbol{h}_{t}=\boldsymbol{h}_{v}+\boldsymbol{h}_{u}/n_i$ to weigh user-side relations' contributions according to sequence lengths. Consequently, as shown in Figure~\ref{fig: model}(b), we devise a position selector and an item selector to select most suitable positions and items to augment a sequence. 

\subsubsection{Position Selector} 
Since the insertion operation needs additional computational costs, inserting items to each position in the target sequence is impractical. To achieve a trade-off between effectiveness and efficiency, our key idea is to select the most suitable positions in the target sequence to insert items instead of randomly or fully selecting positions. From a global-local perspective, when integrating global information into the local sequence, some items may exhibit inconsistent semantics compared to its context in the sequence. The inconsistencies imply that the items at these positions 
are potential noise and that these positions need more information. 

Specifically, we discriminate inconsistencies and select positions by considering two types of detectable signals, \textit{sequentiality} and \textit{similarities}, which individually or jointly reflect whether an item is noisy according to the definition of noisy items in previous denoising studies~\cite{YSS21, QWL21, ZDZ22,LWC23}. An item with high sequentiality means that this item is aligned with its continuous sub-sequence. In contrast, an item with high similarities indicates that it is similar to most items in the target sequence. Therefore, we can detect inconsistencies in the target sequence to select the most possible inconsistent positions to perform sequence augmentation, thus enriching sequence information and alleviating \problem. Next, we will detail the two discriminators used to detect inconsistencies.


{\textbf{Sequentiality Discriminator}.} 
To calculate inconsistency scores (in terms of sequentiality) $\boldsymbol{r}'_{S}\in\mathbb{R}^{n_i}$ of all items in the target sequence $S$ of a length $n_i$, we input $\boldsymbol{H}_S$ into a context-aware encoder as follows
\begin{equation}
\begin{aligned}
    &\boldsymbol{H}^L_S,\boldsymbol{H}^R_S =\text{Bi-LSTM}(\boldsymbol{H}_{S}\vert\Theta_{L},\Theta_{R}),\\
    & \boldsymbol{r}'_{S} = \rho(\sum_{d}(\boldsymbol{H}^L_S\odot\boldsymbol{H}^R_S\odot\boldsymbol{H}_{S})),\\
    & r'_{s_t} = \frac{\text{exp}(\sum_d(\boldsymbol{h}^{L}_{s_t}\odot\boldsymbol{h}^{R}_{s_t}\odot\boldsymbol{h}_{s_t}))}{\sum_{s_t}\text{exp}(\sum_d(\boldsymbol{h}^{L}_{s_t}\odot\boldsymbol{h}^{R}_{s_t}\odot\boldsymbol{h}_{s_t}))},
\end{aligned}
\label{eq: seq discriminator}
\end{equation}
where $\boldsymbol{H}^L_S$ and $\boldsymbol{H}^R_S$ are bi-directional hidden states of the context-aware encoder (Bi-LSTM), $\Theta_{R}$ and $\Theta_{L}$ are the encoder's trainable parameters, $d$ is the dimension, $\rho$ is the Softmax function, and $\odot$ is the element-wise product operator. Here, we utilize a Bi-LSTM as the encoder, as it allows us to learn the sequential dependencies of items in a bidirectional manner. This is crucial for better understanding whether global relations, such as the learned global item transitional relations from $\mathcal{G}$, are inconsistent with local dependencies. Moreover, we consider the strictest conditions (i.e., calculating $\boldsymbol{H}^L_S\odot\boldsymbol{H}^R_S$) to ensure the reliability of the inconsistency scores.

{\textbf{Similarity Discriminator}.} We calculate all items' inconsistency scores $\boldsymbol{r}''_{S}\in\mathbb{R}^{n_i}$ in terms of similarity as follows:
\begin{equation}
\begin{aligned}
        &\boldsymbol{r}''_{S}=\rho(\sum_{n_i}(\boldsymbol{H}_S\boldsymbol{H}^{\mathbf{T}}_S)/(n_i-1)),\\
        &r''_{s_t} = \frac{\text{exp}(\sum_{s_i}(\boldsymbol{h}_{t}\boldsymbol{h}^{\mathsf{T}}_{i})/(n_i-1))}{\sum_{s_t}\text{exp}(\sum_{s_i}(\boldsymbol{h}_{i}\boldsymbol{h}^{\mathsf{T}}_{t})/(n_i-1))}.
\end{aligned}
\label{eq: sim discriminator}
\end{equation}
With the learned sequentiality and similarity scores (i.e., $r'_{s_t}$ and $r''_{s_t}$) of each item $s_t$, we can calculate the likelihood of identifying $s_t$ as inconsistent. Specifically, we can generate a distribution $\boldsymbol{r}_{S}=[r'_{s_1}\cdot r''_{s_1},\cdots,r'_{s_t}\cdot r''_{s_t},\cdots,r'_{s_{n_i}}\cdot r''_{s_{n_i}}]$ representing the potential inconsistencies among items. We pinpoint the most likely inconsistency $s_t$ in $S$ as $s_t = \text{argmax}(\boldsymbol{r}_S)$. However, the $\text{argmax}$ operation is non-differentiable. To identify $s_t$ and facilitate gradient back-propagation, we employ a Gumbel-Softmax function~\cite{WYW19,WLZ19,YGH19,ZLL21} as suggested by previous studies~\cite{QWL21,ZDZ22} to support model learning via
\begin{equation}
\begin{aligned}
    &\boldsymbol{\hat{r}}_{S} = \text{gumbel-softmax}(\boldsymbol{r}_S),\\
    &\hat{r}_{s_t} = \frac{\text{exp}( \log (r_{s_t}) + g_t ) / \tau}{ \sum_{s_i} \text{exp}(\text{log}(r_{s_i}) +g_i )/\tau},
\end{aligned}
\end{equation}
where $\tau > 0$ is the temperature parameter and $g_j$ is i.i.d. sampled from a Gumbel distribution. When $\tau \to 0$, $\boldsymbol{\hat{r}}_{S}$ approximates a one-hot vector (i.e., $\hat{r}_{s_t}=1$ and $\hat{r}_{s_i}=0,\forall s_t\neq s_i\in S$) serving as a hard selection to indicate the most possible inconsistency (i.e., $s_t$) at position $t$ in $S$.
\subsubsection{Item Selector}
Although we can choose a suitable position in the target sequence to insert items, it is challenging to determine which and how many items to select from the entire item universe for insertions. To achieve a trade-off between effectiveness and efficiency, our key idea is to select the two most suitable items. These items should provide sufficient sequential information and can slightly perturb the original sequence. Consequently, we can insert the two items before and after the chosen position in the target sequence. 

Specifically, we use the context-aware encoder to match suitable items from the entire item universe. Following the selection of positions, the encoder can effectively detect inconsistent insertions to avoid introducing additional noise. For the chosen position (e.g., $s_t$'s position $t$ in the target sequence), we encode its bi-directional contextual information using the context-aware encoder and rank the two most suitable items from the entire item universe, taking into account $s_t$'s sequential dependencies. Mathematically, we select two items (embeddings) $\boldsymbol{h}^{L}_{t},\boldsymbol{h}^{R}_{t}\in\mathbb{R}^d$ as follows 
\begin{equation}
\begin{aligned}
    \boldsymbol{q}^{L},\boldsymbol{q}^{R} = \text{Bi-LSTM}&(\boldsymbol{H}_{S}\vert\Theta_{L},\Theta_{R})_{s_t},\\
    \boldsymbol{k}^{L} = \boldsymbol{q}^{L}\boldsymbol{H}_{v}&; \boldsymbol{k}^{R} = \boldsymbol{q}^{R}\boldsymbol{H}_{v},\\
    \boldsymbol{\hat{k}}^{L} \!\!\!=\!\!\! \text{gumbel-softmax}(\boldsymbol{k}^{L})&;\boldsymbol{\hat{k}}^{R}\!\!\!=\!\!\!\text{gumbel-softmax}(\boldsymbol{k}^{R}),\\
    \boldsymbol{h}^{L}_{t} = \sum_{\vert\mathcal{V}\vert}(\boldsymbol{\hat{k}}^{L}\boldsymbol{H}_{v})&;\boldsymbol{h}^{R}_{t} = \sum_{\vert\mathcal{V}\vert}(\boldsymbol{\hat{k}}^{R}\boldsymbol{H}_{v}),
\end{aligned}
\end{equation}
where $\boldsymbol{H}_{v}$ represents all items' embeddings and $\boldsymbol{\hat{k}}^{L},\boldsymbol{\hat{k}}^{R}\in\mathbb{R}^{\vert\mathcal{V}\vert}$ are hard-coding binary vectors. Only one element of $\boldsymbol{\hat{k}}^{L}$ ($\boldsymbol{\hat{k}}^{R}$) is 1 and the rest are 0. Accordingly, we can lengthen the sequence by inserting the two items into $\boldsymbol{H}_{S}$ to generate augmented representation sequence as $\boldsymbol{H}'_{S}=[\boldsymbol{h}_{s_1},\cdots,\boldsymbol{h}^{L}_{t},\boldsymbol{h}_{t},\boldsymbol{h}^{R}_{t},\cdots,\boldsymbol{h}_{s_{n_i}}]$ to alleviate \problem.

\subsection{Hierarchical Denoising Module}
To further eliminate inappropriate augmentations that might introduce additionally noisy items in the second stage of \model, we propose a hierarchical denoising module, as illustrated in Figure~\ref{fig: model}(c). This module can gradually refine the augmented sequence and remove all potentially noisy items. Formally, we can generate a final augmented item representation sequence $\boldsymbol{H}''_{S}$ for any sequence denoising method as
\begin{equation}
    \boldsymbol{H}''_{S} = f_{hdm}(\boldsymbol{H}_{S},\boldsymbol{H}'_{S},\Theta_{hdm}),
\end{equation}
where $f_{hdm}$ is the same position selector used in Eq.~(\ref{eq: seq discriminator}) and Eq.~(\ref{eq: sim discriminator}). Consequently, we can use any denoising model $f_{den}$ (e.g., HSD~\cite{ZDZ22}) to generate the noiseless sub-sequence:
\begin{equation}
    \boldsymbol{H}^{-}_{S} = f_{den}(\boldsymbol{H}_{S}\vert\boldsymbol{H}''_{S},\Theta_{den}),
\label{eq: denoise}
\end{equation}
where $\Theta_{den}$ is the set of parameters of $f_{den}$. We highlight that the proposed self-augmentation method can serve as a plug-in to enhance the reliability of any sequence denoising method (based on intra-sequence information) by alleviating \problem.
\subsection{Sequential Recommender}
As illustrated in Figure~\ref{fig: graphs}(c), with the generated noiseless item embedding sequence $\boldsymbol{H}^{-}_{S}$, we can feed $\boldsymbol{H}^{-}_{S}$ into various sequential recommendation models via
\begin{equation}
    \boldsymbol{h}_{S^i} = f_{seq}(\boldsymbol{H}^{-}_{S}\vert\Theta_{seq}),
\end{equation}
where $f_{seq}$ is any existing sequential recommendation models (e.g., SASRec~\cite{KM18}), and $\Theta_{seq}$ is the set of parameters of $f_{seq}$. Consequently, we calculate the similarity score (i.e., dot-product operation) between the sequence representation $\boldsymbol{h}_{S^i}$ and the entire item universe to make recommendations, i.e., recommending the top-$k$ items. Note that we only adopt the second stage in the training process instead of validating or testing, i.e., $\boldsymbol{H}'_{s}=f_{den}(\boldsymbol{H}_{s}\vert\Theta_{den})$ in Eq.~(\ref{eq: denoise}). This decision is made because the self-augmentation module is designed to lengthen short sequences in order to tackle \problem. Thus, after training, the jointly learned denoising model is capable of generating reliable denoising results.

\subsection{Model Complexity Analysis}
\subsubsection{Space Complexity Analysis} We analyze the size of the trainable parameters in each stage of \model. 

In this first stage, the parameters of the embedding layer and the global relation encoder are denoted by $\Theta_1$. We have $\vert\Theta_1\vert =(\vert\mathcal{V}\vert+\vert\mathcal{U}\vert)\times d+\vert \Theta_{g}\vert$, where $\Theta_g$ means the set of parameters used in the global relation encoder. In the second stage, we denote the self-augmentation module's parameters by $\Theta_2$. Its size $\vert\Theta_2\vert=\vert\Theta_{L}\vert=\vert\Theta_{R}\vert$. In the third stage, we generate a noiseless sequence and make recommendations by a hierarchical denoising module and a sequential recommender. We denote the parameters by $\Theta_3$, and we have $\vert\Theta_{3}\vert=\vert\Theta_{den}\vert+\vert\Theta_{seq}\vert$. 
In conclusion, since $\mathcal{V}$ and $\mathcal{U}$ are typically of a large size (i.e., $\vert\mathcal{V}\vert+\vert\mathcal{U}\vert\gg d$ and $\vert\mathcal{V}\vert\gg n$.), the proposed \model can achieve comparable space complexity with sequential recommendation models (e.g., SASRec~\cite{KM18}) and the state-of-the-art sequence denoising models (e.g., HSD~\cite{ZDZ22}) as $\boldsymbol{O}(\vert\mathcal{V}\vert+\vert\mathcal{U}\vert)$.

\begin{table}[t]
\caption{Statistics of Experimental Datasets.}
\label{tab: dataset}
\resizebox{\linewidth}{!}{\begin{tabular}{lrrrrc}
\toprule
\multicolumn{1}{c}{\textbf{Dataset}} & \textbf{\# Users} & \textbf{\# Items} & \textbf{\# Actions} & \textbf{\# Avg. lens} & \textbf{\# Sparsity} \\ \midrule
Beauty                               & 22,364            & 12,102            & 198,502             & 8.9                    & 99.93\%              \\
Sports                               & 35,599            & 18,358            & 296,337             & 8.3                    & 99.95\%              \\
Yelp                                 & 30,495            & 20,062            & 317,078             & 10.4                   & 99.95\%              \\
ML-100K                              & 944               & 1,350             & 99,287              & 105.3                  & 92.21\%              \\
ML-1M                                & 6,041             & 3,417             & 999,611             & 165.5                  & 95.16\%              \\ \bottomrule
\end{tabular}}
\vspace{-3mm}
\end{table}

\begin{table*}[t]
\caption{Experimental results of different sequential recommenders, with (w) or without (w/o) integrating with SSDRec, are Reported. The best results are marked as bold. All Improvements Are Statistically Significant (i.e., Two-sided t-tests with $p<0.05$)}
\label{tab:Overall-comprehension-1}
\centering
\resizebox{\textwidth}{!}{%
\begin{tabular}{clcccccccccccc}
\toprule
\multirow{2}{*}{\textbf{DataSet}} &
  \multicolumn{1}{c}{\multirow{2}{*}{\textbf{Metric}}} &
  \multicolumn{2}{c}{\textbf{GRU4Rec}} &
  \multicolumn{2}{c}{\textbf{NARM}} &
  \multicolumn{2}{c}{\textbf{STAMP}} &
  \multicolumn{2}{c}{\textbf{Caser}} &
  \multicolumn{2}{c}{\textbf{SASRec}} &
  \multicolumn{2}{c}{\textbf{BERT4Rec}} \\ \cmidrule{3-14} 
 &
  \multicolumn{1}{c}{} &
  w/o &
  w &
  w/o &
  w &
  w/o &
  w &
  w/o &
  w &
  w/o &
  w &
  w/o &
  w \\ \midrule
\multirow{8}{*}{ML-100K} &
  HR@5 &
  0.0191 &
  \textbf{0.0276} &
  0.0180 &
  \textbf{0.0212} &
  0.0201 &
  \textbf{0.0201} &
  0.0212 &
  \textbf{0.0286} &
  0.0191 &
  \textbf{0.0456} &
  0.0191 &
  \textbf{0.0329} \\
 &
  HR@10 &
  0.0286 &
  \textbf{0.0520} &
  0.0403 &
  \textbf{0.0422} &
  0.0392 &
  \textbf{0.0467} &
  0.0339 &
  \textbf{0.0530} &
  0.0371 &
  \textbf{0.0795} &
  0.0414 &
  \textbf{0.0700} \\
 &
  HR@20 &
  0.0594 &
  \textbf{0.0721} &
  0.0657 &
  \textbf{0.0827} &
  0.0700 &
  \textbf{0.0817} &
  0.0679 &
  \textbf{0.1145} &
  0.0764 &
  \textbf{0.1336} &
  0.0912 &
  \textbf{0.1230} \\
 &
  N@5 &
  0.0104 &
  \textbf{0.0159} &
  0.0132 &
  \textbf{0.0132} &
  0.0115 &
  \textbf{0.0115} &
  0.0113 &
  \textbf{0.0177} &
  0.0114 &
  \textbf{0.0262} &
  0.0117 &
  \textbf{0.0176} \\
 &
  N@10 &
  0.0134 &
  \textbf{0.0237} &
  0.0202 &
  \textbf{0.0207} &
  0.0176 &
  \textbf{0.0199} &
  0.0153 &
  \textbf{0.0256} &
  0.0172 &
  \textbf{0.0379} &
  0.0189 &
  \textbf{0.0297} \\
 &
  N@20 &
  0.0212 &
  \textbf{0.0288} &
  0.0267 &
  \textbf{0.0300} &
  0.0253 &
  \textbf{0.0287} &
  0.0238 &
  \textbf{0.0409} &
  0.0270 &
  \textbf{0.0507} &
  0.0315 &
  \textbf{0.0431} \\
 &
  MRR &
  0.0109 &
  \textbf{0.0167} &
  0.0162 &
  \textbf{0.0162} &
  0.0132 &
  \textbf{0.0145} &
  0.0119 &
  \textbf{0.0214} &
  0.0139 &
  \textbf{0.0290} &
  0.0157 &
  \textbf{0.0213} \\ \cmidrule{2-14} 
 &
  Improvement &
   &
  \textbf{52.36\%} &
  \textbf{} &
  \textbf{9.03\%} &
   &
  \textbf{10.31\%} &
   &
  \textbf{62.22\%} &
   &
  \textbf{110.64\%} &
   &
  \textbf{50.90\%} \\ \midrule
\multirow{8}{*}{ML-1M} &
  HR@5 &
  0.0207 &
  \textbf{0.0308} &
  0.0151 &
  \textbf{0.0255} &
  0.0104 &
  \textbf{0.0248} &
  0.0232 &
  \textbf{0.0349} &
  0.0397 &
  \textbf{0.0492} &
  0.0224 &
  \textbf{0.0429} \\
 &
  HR@10 &
  0.0359 &
  \textbf{0.0584} &
  0.0349 &
  \textbf{0.0437} &
  0.0215 &
  \textbf{0.0445} &
  0.0440 &
  \textbf{0.0639} &
  0.0666 &
  \textbf{0.0896} &
  0.0495 &
  \textbf{0.0778} \\
 &
  HR@20 &
  0.0613 &
  \textbf{0.0942} &
  0.0591 &
  \textbf{0.0684} &
  0.0589 &
  \textbf{0.0715} &
  0.0677 &
  \textbf{0.1036} &
  0.1007 &
  \textbf{0.1412} &
  0.0980 &
  \textbf{0.1267} \\
 &
  N@5 &
  0.0130 &
  \textbf{0.0188} &
  0.0080 &
  \textbf{0.0156} &
  0.0063 &
  \textbf{0.0145} &
  0.0150 &
  \textbf{0.0219} &
  0.0207 &
  \textbf{0.0292} &
  0.0132 &
  \textbf{0.0249} \\
 &
  N@10 &
  0.0178 &
  \textbf{0.0278} &
  0.0144 &
  \textbf{0.0214} &
  0.0099 &
  \textbf{0.0209} &
  0.0218 &
  \textbf{0.0313} &
  0.0294 &
  \textbf{0.0410} &
  0.0218 &
  \textbf{0.0361} \\
 &
  N@20 &
  0.0242 &
  \textbf{0.0367} &
  0.0205 &
  \textbf{0.0276} &
  0.0194 &
  \textbf{0.0277} &
  0.0278 &
  \textbf{0.0412} &
  0.0379 &
  \textbf{0.0544} &
  0.0339 &
  \textbf{0.0484} \\
 &
  MRR &
  0.0142 &
  \textbf{0.0210} &
  0.0100 &
  \textbf{0.0164} &
  0.0091 &
  \textbf{0.0156} &
  0.0168 &
  \textbf{0.0242} &
  0.0203 &
  \textbf{0.0315} &
  0.0169 &
  \textbf{0.0270} \\ \cmidrule{2-14} 
 &
  Improvement &
   &
  \textbf{52.21\%} &
   &
  \textbf{50.30\%} &
   &
  \textbf{88.90\%} &
   &
  \textbf{47.22\%} &
   &
  \textbf{39.70\%} &
   &
  \textbf{62.11\%} \\ \midrule
\multirow{8}{*}{Beauty} &
  HR@5 &
  0.0077 &
  \textbf{0.0218} &
  0.0120 &
  \textbf{0.0205} &
  0.0080 &
  \textbf{0.0223} &
  0.0072 &
  \textbf{0.0228} &
  0.0232 &
  \textbf{0.0313} &
  0.0060 &
  \textbf{0.0279} \\
 &
  HR@10 &
  0.0135 &
  \textbf{0.0358} &
  0.0208 &
  \textbf{0.0339} &
  0.0135 &
  \textbf{0.0366} &
  0.0130 &
  \textbf{0.0353} &
  0.0386 &
  \textbf{0.0517} &
  0.0127 &
  \textbf{0.0468} \\
 &
  HR@20 &
  0.0256 &
  \textbf{0.0567} &
  0.0366 &
  \textbf{0.0530} &
  0.0231 &
  \textbf{0.0570} &
  0.0237 &
  \textbf{0.0527} &
  0.0583 &
  \textbf{0.0776} &
  0.0204 &
  \textbf{0.0716} \\
 &
  N@5 &
  0.0045 &
  \textbf{0.0136} &
  0.0071 &
  \textbf{0.0130} &
  0.0046 &
  \textbf{0.0141} &
  0.0043 &
  \textbf{0.0146} &
  0.0122 &
  \textbf{0.0180} &
  0.0037 &
  \textbf{0.0157} \\
 &
  N@10 &
  0.0064 &
  \textbf{0.0181} &
  0.0099 &
  \textbf{0.0173} &
  0.0064 &
  \textbf{0.0187} &
  0.0062 &
  \textbf{0.0186} &
  0.0172 &
  \textbf{0.0241} &
  0.0059 &
  \textbf{0.0218} \\
 &
  N@20 &
  0.0094 &
  \textbf{0.0233} &
  0.0139 &
  \textbf{0.0221} &
  0.0088 &
  \textbf{0.0238} &
  0.0089 &
  \textbf{0.0229} &
  0.0222 &
  \textbf{0.0306} &
  0.0078 &
  \textbf{0.0281} \\
 &
  MRR &
  0.0051 &
  \textbf{0.0141} &
  0.0077 &
  \textbf{0.0136} &
  0.0049 &
  \textbf{0.0147} &
  0.0049 &
  \textbf{0.0147} &
  0.0121 &
  \textbf{0.0179} &
  0.0044 &
  \textbf{0.0160} \\ \cmidrule{2-14} 
 &
  Improvement &
   &
  \textbf{168.45\%} &
   &
  \textbf{67.44\%} &
   &
  \textbf{180.83\%} &
   &
  \textbf{186.77\%} &
   &
  \textbf{39.34\%} &
   &
  \textbf{286.03\%} \\ \midrule
\multirow{8}{*}{Sports} &
  HR@5 &
  0.0064 &
  \textbf{0.0113} &
  0.0090 &
  \textbf{0.0101} &
  0.0071 &
  \textbf{0.0128} &
  0.0069 &
  \textbf{0.0106} &
  0.0111 &
  \textbf{0.0165} &
  0.0055 &
  \textbf{0.0153} \\
 &
  HR@10 &
  0.0114 &
  \textbf{0.0184} &
  0.0138 &
  \textbf{0.0166} &
  0.0123 &
  \textbf{0.0194} &
  0.0115 &
  \textbf{0.0175} &
  0.0177 &
  \textbf{0.0259} &
  0.0104 &
  \textbf{0.0261} \\
 &
  HR@20 &
  0.0183 &
  \textbf{0.0295} &
  0.0223 &
  \textbf{0.0274} &
  0.0182 &
  \textbf{0.0293} &
  0.0178 &
  \textbf{0.0274} &
  0.0270 &
  \textbf{0.0391} &
  0.0167 &
  \textbf{0.0404} \\
 &
  N@5 &
  0.0035 &
  \textbf{0.0073} &
  0.0058 &
  \textbf{0.0063} &
  0.0046 &
  \textbf{0.0085} &
  0.0046 &
  \textbf{0.0069} &
  0.0058 &
  \textbf{0.0092} &
  0.0036 &
  \textbf{0.0082} \\
 &
  N@10 &
  0.0051 &
  \textbf{0.0095} &
  0.0073 &
  \textbf{0.0084} &
  0.0062 &
  \textbf{0.0106} &
  0.0061 &
  \textbf{0.0091} &
  0.0079 &
  \textbf{0.0122} &
  0.0051 &
  \textbf{0.0116} \\
 &
  N@20 &
  0.0068 &
  \textbf{0.0123} &
  0.0094 &
  \textbf{0.0111} &
  0.0077 &
  \textbf{0.0131} &
  0.0077 &
  \textbf{0.0116} &
  0.0103 &
  \textbf{0.0156} &
  0.0067 &
  \textbf{0.0152} \\
 &
  MRR &
  0.0036 &
  \textbf{0.0076} &
  0.0059 &
  \textbf{0.0066} &
  0.0048 &
  \textbf{0.0086} &
  0.0049 &
  \textbf{0.0072} &
  0.0056 &
  \textbf{0.0089} &
  0.0040 &
  \textbf{0.0082} \\ \cmidrule{2-14} 
 &
  Improvement &
   &
  \textbf{83.72\%} &
   &
  \textbf{15.57\%} &
  \textbf{} &
  \textbf{72.01\%} &
   &
  \textbf{50.93\%} &
   &
  \textbf{51.89\%} &
   &
  \textbf{136.88\%} \\ \midrule
\multirow{8}{*}{Yelp} &
  HR@5 &
  0.0045 &
  \textbf{0.0163} &
  0.0089 &
  \textbf{0.0155} &
  0.0053 &
  \textbf{0.0153} &
  0.0054 &
  \textbf{0.0137} &
  0.0290 &
  \textbf{0.0358} &
  0.0083 &
  \textbf{0.0258} \\
 &
  HR@10 &
  0.0083 &
  \textbf{0.0218} &
  0.0152 &
  \textbf{0.0225} &
  0.0092 &
  \textbf{0.0235} &
  0.0083 &
  \textbf{0.0226} &
  0.0349 &
  \textbf{0.0479} &
  0.0164 &
  \textbf{0.0353} \\
 &
  HR@20 &
  0.0152 &
  \textbf{0.0310} &
  0.0267 &
  \textbf{0.0344} &
  0.0156 &
  \textbf{0.0376} &
  0.0127 &
  \textbf{0.0378} &
  0.0437 &
  \textbf{0.0656} &
  0.0291 &
  \textbf{0.0518} \\
 &
  N@5 &
  0.0028 &
  \textbf{0.0127} &
  0.0056 &
  \textbf{0.0112} &
  0.0037 &
  \textbf{0.0102} &
  0.0037 &
  \textbf{0.0091} &
  0.0249 &
  \textbf{0.0270} &
  0.0050 &
  \textbf{0.0200} \\
 &
  N@10 &
  0.0040 &
  \textbf{0.0145} &
  0.0076 &
  \textbf{0.0135} &
  0.0049 &
  \textbf{0.0128} &
  0.0047 &
  \textbf{0.0120} &
  0.0268 &
  \textbf{0.0309} &
  0.0076 &
  \textbf{0.0230} \\
 &
  N@20 &
  0.0057 &
  \textbf{0.0168} &
  0.0105 &
  \textbf{0.0165} &
  0.0065 &
  \textbf{0.0164} &
  0.0057 &
  \textbf{0.0158} &
  0.0291 &
  \textbf{0.0354} &
  0.0108 &
  \textbf{0.0271} \\
 &
  MRR &
  0.0031 &
  \textbf{0.0129} &
  0.0061 &
  \textbf{0.0115} &
  0.0041 &
  \textbf{0.0106} &
  0.0038 &
  \textbf{0.0098} &
  0.0250 &
  \textbf{0.0269} &
  0.0058 &
  \textbf{0.0204} \\ \cmidrule{2-14} 
 &
  Improvement &
   &
  \textbf{236.54\%} &
  \textbf{} &
  \textbf{67.76\%} &
  \textbf{} &
  \textbf{161.84\%} &
  \textbf{} &
  \textbf{165.71\%} &
  \textbf{} &
  \textbf{23.40\%} &
  \textbf{} &
  \textbf{187.05\%} \\ \bottomrule
\vspace{-3mm}
\end{tabular}%
}
\end{table*}

\begin{table*}[ht]
\caption{The Experimental Comparison between \model and the State-of-the-art Denoising Methods. The Best Results Are Bold, and the Second-best Results are Underlined. All Improvements Are Statistically Significant (i.e., Two-sided t-test with $p<0.05$).}
\label{tab:Overall-comprehension-2}
\centering
\begin{tabular}{clccccccc}
\toprule
\textbf{Dataset} &
  \multicolumn{1}{c}{\textbf{Model}} &
  HR@5 &
  HR@10 &
  HR@20 &
  N@5 &
  N@10 &
  N@20 &
  MRR \\ \midrule
\multirow{7}{*}{ML-100K} &
  \textbf{DSAN} &
  0.0201 &
  0.0435 &
  0.0700 &
  0.0115 &
  0.0188 &
  0.0254 &
  0.0133 \\
 &
  \textbf{FMLP-Rec} &
  0.0170 &
  0.0477 &
  0.0764 &
  0.0117 &
  0.0216 &
  0.0288 &
  0.0160 \\
 &
  \textbf{HSD} &
  0.0370 &
  0.0583 &
  0.1018 &
  {\ul 0.0225} &
  0.0292 &
  0.0401 &
  0.0234 \\
 &
  \textbf{DCRec} &
  0.0244 &
  0.0424 &
  0.0806 &
  0.0156 &
  0.0214 &
  0.0308 &
  0.0175 \\
 &
  \textbf{STEAM} &
  {\ul 0.0371} &
  {\ul 0.0647} &
  {\ul 0.1220} &
  0.0220 &
  {\ul 0.0309} &
  {\ul 0.0452} &
  {\ul 0.0245} \\ \cmidrule{2-9} 
 &
  \textbf{SSDRec} &
  \textbf{0.0456} &
  \multicolumn{1}{l}{\textbf{0.0795}} &
  \textbf{0.1336} &
  \multicolumn{1}{l}{\textbf{0.0262}} &
  \multicolumn{1}{l}{\textbf{0.0379}} &
  \multicolumn{1}{l}{\textbf{0.0507}} &
  \multicolumn{1}{l}{\textbf{0.0290}} \\ \cmidrule{2-9} 
 &
  Improvement &
  \textbf{22.91\%} &
  \textbf{22.87\%} &
  \textbf{9.51\%} &
  \textbf{16.44\%} &
  \textbf{22.65\%} &
  \textbf{12.17\%} &
  \textbf{18.37\%} \\ \midrule
\multirow{7}{*}{ML-1M} &
  \textbf{DSAN} &
  0.0098 &
  0.0336 &
  0.0651 &
  0.0048 &
  0.0122 &
  0.0200 &
  0.0081 \\
 &
  \textbf{FMLP-Rec} &
  0.0210 &
  0.0449 &
  0.0707 &
  0.0120 &
  0.0199 &
  0.0263 &
  0.0142 \\
 &
  \textbf{HSD} &
  0.0440 &
  {\ul 0.0823} &
  {\ul 0.1326} &
  0.0251 &
  {\ul 0.0373} &
  {\ul 0.0515} &
  {\ul 0.0278} \\
 &
  \textbf{DCRec} &
  {\ul 0.0440} &
  0.0753 &
  0.1111 &
  {\ul 0.0251} &
  0.0351 &
  0.0442 &
  0.0254 \\
 &
  \textbf{STEAM} &
  0.0127 &
  0.0270 &
  0.0548 &
  0.0060 &
  0.0105 &
  0.0171 &
  0.0073 \\ \cmidrule{2-9} 
 &
  \textbf{SSDRec} &
  \textbf{0.0492} &
  \textbf{0.0896} &
  \textbf{0.1412} &
  \textbf{0.0292} &
  \textbf{0.0410} &
  \textbf{0.0544} &
  \textbf{0.0315} \\ \cmidrule{2-9} 
 &
  Improvement &
  \textbf{11.82\%} &
  \textbf{8.87\%} &
  \textbf{6.49\%} &
  \textbf{16.33\%} &
  \textbf{9.92\%} &
  \textbf{5.63\%} &
  \textbf{13.31\%} \\ \midrule
\multirow{7}{*}{Beauty} &
  \textbf{DSAN} &
  0.0092 &
  0.0152 &
  0.0264 &
  0.0058 &
  0.0077 &
  0.0105 &
  0.0062 \\
 &
  \textbf{FMLP-Rec} &
  0.0095 &
  0.0166 &
  0.0284 &
  0.0056 &
  0.0078 &
  0.0107 &
  0.0060 \\
 &
  \textbf{HSD} &
  0.0247 &
  0.0435 &
  0.0674 &
  0.0137 &
  0.0197 &
  0.0260 &
  0.0142 \\
 &
  \textbf{DCRec} &
  {\ul 0.0312} &
  {\ul 0.0499} &
  {\ul 0.0714} &
  {\ul 0.0177} &
  {\ul 0.0234} &
  {\ul 0.0288} &
  {\ul 0.0167} \\
 &
  \textbf{STEAM} &
  0.0141 &
  0.0295 &
  0.0521 &
  0.0074 &
  0.0124 &
  0.0180 &
  0.0088 \\ \cmidrule{2-9} 
 &
  \textbf{SSDRec} &
  \textbf{0.0313} &
  \textbf{0.0517} &
  \textbf{0.0776} &
  \textbf{0.0180} &
  \textbf{0.0241} &
  \textbf{0.0306} &
  \textbf{0.0179} \\ \cmidrule{2-9} 
 &
  Improvement &
  \textbf{0.32\%} &
  \textbf{3.61\%} &
  \textbf{8.68\%} &
  \textbf{1.69\%} &
  \textbf{2.99\%} &
  \textbf{6.25\%} &
  \textbf{7.19\%} \\ \midrule
\multirow{7}{*}{Sports} &
  \textbf{DSAN} &
  0.0061 &
  0.0105 &
  0.0215 &
  0.0042 &
  0.0056 &
  0.0084 &
  0.0049 \\
 &
  \textbf{FMLP-Rec} &
  0.0068 &
  0.0117 &
  0.0180 &
  0.0044 &
  0.0059 &
  0.0075 &
  0.0046 \\
 &
  \textbf{HSD} &
  0.0131 &
  0.0215 &
  0.0337 &
  0.0071 &
  0.0100 &
  0.0132 &
  0.0071 \\
 &
  \textbf{DCRec} &
  {\ul 0.0160} &
  {\ul 0.0255} &
  {\ul 0.0377} &
  {\ul 0.0090} &
  {\ul 0.0121} &
  {\ul 0.0146} &
  {\ul 0.0084} \\
 &
  \textbf{STEAM} &
  0.0063 &
  0.0133 &
  0.0251 &
  0.0031 &
  0.0053 &
  0.0083 &
  0.0038 \\ \cmidrule{2-9} 
 &
  \textbf{SSDRec} &
  \textbf{0.0165} &
  \textbf{0.0259} &
  \textbf{0.0391} &
  \textbf{0.0092} &
  \textbf{0.0122} &
  \textbf{0.0156} &
  \textbf{0.0089} \\ \cmidrule{2-9} 
 &
  Improvement &
  \textbf{3.13\%} &
  \textbf{1.57\%} &
  \textbf{3.71\%} &
  \textbf{2.22\%} &
  \textbf{0.83\%} &
  \textbf{6.85\%} &
  \textbf{5.95\%} \\ \midrule
\multirow{7}{*}{Yelp} &
  \textbf{DSAN} &
  0.0269 &
  0.0369 &
  0.0541 &
  0.0211 &
  0.0242 &
  0.0285 &
  0.0216 \\
 &
  \textbf{FMLP-Rec} &
  0.0203 &
  0.0294 &
  0.0436 &
  0.0142 &
  0.0171 &
  0.0207 &
  0.0144 \\
 &
  \textbf{HSD} &
  0.0296 &
  0.0388 &
  0.0557 &
  0.0231 &
  0.0260 &
  0.0303 &
  0.0233 \\
 &
  \textbf{DCRec} &
  {\ul 0.0316} &
  {\ul 0.0433} &
  {\ul 0.0611} &
  {\ul 0.0242} &
  {\ul 0.0280} &
  {\ul 0.0324} &
  {\ul 0.0246} \\
 &
  \textbf{STEAM} &
  0.0255 &
  0.0389 &
  0.0590 &
  0.0168 &
  0.0211 &
  0.0262 &
  0.0171 \\ \cmidrule{2-9} 
 &
  \textbf{SSDRec} &
  \textbf{0.0358} &
  \textbf{0.0479} &
  \textbf{0.0656} &
  \textbf{0.0270} &
  \textbf{0.0309} &
  \textbf{0.0354} &
  \textbf{0.0269} \\ \cmidrule{2-9} 
 & Improvement &
  \textbf{13.29\%} &
  \textbf{10.62\%} &
  \textbf{7.36\%} &
  \textbf{11.57\%} &
  \textbf{10.36\%} &
  \textbf{9.26\%} &
  \textbf{9.35\%} \\ \bottomrule
\end{tabular}%
\end{table*}

\subsubsection{Time Complexity Analysis}
We analyze the time complexity of relation learning and sequence augmentation. Specifically, \model takes $\boldsymbol{O}(\text{max}(\vert\mathcal{U}\vert,\vert\mathcal{V}\vert)^2)$ time complexity for learning $\mathcal{G}$ in the first stage, and spends $\boldsymbol{O}(B\times n)$, $\boldsymbol{O}(B\times\vert\mathcal{V}\vert)$, and $\boldsymbol{O}(B\times n)$ for selecting positions, items, and inserting items, respectively, where $B$ is the batch size. In conclusion, since $\vert\mathcal{V}\vert\gg n$ for short sequences and the batch size is a constant, \model takes $\boldsymbol{O}(\text{max}(\vert\mathcal{U}\vert,\vert\mathcal{V}\vert)^2)$ additional time complexity, which is acceptable in practice, compared to the existing  sequence denoising methods (e.g., HSD~\cite{ZDZ22} and STEAM~\cite{LWC23}) for fully sorting items to make recommendations,  whose complexity is $\boldsymbol{O}(\vert\mathcal{U}\vert \times \vert\mathcal{V}\vert)$. 
Since the five types of relations in the global relation encoder can be processed independently, we can further speed up the training process by utilizing parallel computing techniques.

\section{Evaluation}
\label{sec:experiments}

In this section, we evaluate our proposed \model method and answer the following key research questions:
\begin{itemize}[\IEEElabelindent=0pt]
    \item \textbf{RQ1}: Does the proposed \model framework demonstrate flexible applicability and effectiveness across various mainstream sequential recommendation methods?
    \item \textbf{RQ2}: Does the proposed \model model outperform state-of-the-art denoising methods?
    \item \textbf{RQ3}: How do different stages of \model contribute to the performance of sequential recommendation?
    \item \textbf{RQ4}: How efficient is \model compared with baselines?
    \item \textbf{RQ5}: How is the interpretation ability of the three-stage sequence denoising for sequential recommendation?
    \item \textbf{RQ6}: How sensitive is \model to hyperparameters?
\end{itemize}

\subsection{Experimental Settings}
\subsubsection{Datasets and Evaluation
Metrics} To evaluate the effectiveness of \model under different sequence lengths, we conduct experiments on five public recommendation datasets: 
(1) \textbf{MovieLens\footnote{https://movielens.org/}} that contains users' reviews and ratings on movies. We use the 100K and 1M versions (ML-100K and ML-1M for short) in the experiments. 
(2) \textbf{Amazon-Beauty and Sports\footnote{http://jmcauley.ucsd.edu/data/amazon/}} that is collected from the Amazon platform, recording users' historical purchases over the abundant products. We conduct experiments on two representative subcategories: beauty and sports. 
(3) \textbf{Yelp\footnote{https://www.yelp.com/dataset}} that is a large-scale and popular dataset for business recommendations. It contains user reviews of restaurants and bars. Due to its large size, we only use the transaction records after January 1st, 2019. Following the previous studies~\cite{SLW19,YSS21, ZYZ22, ZDZ22}, we filter out short sequences with less than 5 items and infrequent items whose frequency is less than 5. We set the maximum sequence length to 200 for ML-1M and 50 for other datasets. Moreover, we adopt the widely used \textit{leave-one-out} strategy by leaving users’ last interacted items in the sequences as the training, valid, and test sets. We consider three evaluation metrics, Hit Ratio (HR@K), Normalized Discounted Cumulative Gain (NDCG@K), and Mean Reciprocal Rank (MRR@K), for performance evaluation. Specifically, we report results on HR@\{5,10,20\} (N@K for short), NDCG@\{5,10,20\} and MRR@20 (MRR for short) over the entire item universe (i.e., full ranking), instead of sampling, to avoid \textcolor{black}{the bias introduced by the sampling process}~\cite{KR20,CWS21}. The statistics of the five datasets are summarized in Table~\ref{tab: dataset}.

\subsubsection{Baselines}
We compare \model with various representative recommendation methods, including conventional sequential recommendation and sequence denoising models. The conventional sequential recommendation models include: 
\begin{itemize}[\IEEElabelindent=0pt]
    \item \textbf{GRU4Rec}~\cite{HKB15} uses gated recurrent units (GRUs) and incorporates a ranking-based loss to learn representations.
    \item \textbf{Caser}~\cite{TW18} utilizes two horizontal and vertical convolutions to learn sequence representations.
    \item \textbf{NARM}~\cite{LRC17} leverages a hybrid encoder with an attention mechanism to model user sequential interaction patterns.
    \item \textbf{STAMP}~\cite{LZM18} is empowered by a short-term attention/memory priority model for user intent learning.
    \item \textbf{SASRec}~\cite{KM18} is a pioneer in utilizing the multi-head attention mechanism to learn sequence representations.
    \item \textbf{BERT4Rec}~\cite{SLW19} introduces a reconstruction task and leverages the deep bidirectional Transformer to learn informative sequential dependencies in the context of a sequence. 
\end{itemize}

In addition, we compare \model with five state-of-the-art sequence denoising methods, including both implicit and explicit sequence denoising models: 
\begin{itemize}[\IEEElabelindent=0pt]

    \item \textbf{FMLP-Rec}~\cite{ZYZ22}, which is an implicit sequence denoising model, utilizes a Fast Fourier Transform (FFT) and an inverse FFT procedure to reduce noise's negative influence. We use the filter-enhanced MLP as FMLP-Rec's base model to learn sequence representations. 
    \item \textbf{DSAN}~\cite{YSS21}, which is an explicit sequence denoising model, explores a sparse Transformer function to model the correlations between each item and a virtual target item to explicitly identify noise. In the experiments, we take the dual attention network as DSAN's base model. 
    \item \textbf{HSD}~\cite{ZDZ22}, which is an explicit sequence denoising model, pinpoints noise by additionally considering to detect items that are inconsistent in terms of sequentiality and user interests. Following the original paper, we take BERT4Rec~\cite{SLW19} as HSD's backbone to conduct comparison experiments.
    \item \textbf{STEAM}~\cite{LWC23}, which is an explicit sequence denoising model, randomly inserts and deletes items in raw sequences to enforce the proposed corrector to reconstruct original sequences for denoising. To conduct comparison experiments, we take the bi-directional Transformer used in the original paper as STEAM's base model.
    \item \textbf{DCRec}~\cite{YHX23}, which is a debiased model, uses a debiased contrastive learning method to capture item-transition patterns in sequences and dependencies between users to facilitate sequential pattern encoding. We consider the multi-channel conformity weighting network and Transformer layer in the experiments.
\end{itemize}
We do not compare SSDRec with a few recent denoising methods~\cite{GDH22,ZCZH23,LZW23} because their settings are different from the setting of sequence denoising in sequential recommendation.

\subsubsection{Implement Details} 
Identical to the settings of previous methods~\cite{LRC17,LZM18,WTZ19,YSS21,HZC22,ZDZ22}, we set the default embedding size $d$ to 100 and mini-batch size to 256 for all methods, and the embedding parameters are initialized with Xavier~\cite{GB10}. We use the Adam optimizer~\cite{KB14} with a default learning rate of 0.001, and adopt the early-stopping training strategy \textcolor{black}{if the HR@20 performance on the validation set decreases for 10 continuous epochs}. The $L_2$ regularization coefficient is searched in $\{0, 10^{-3}, 10^{-4}\}$. In particular, we set the ratio of few-shot users and items as 0.9 and 0.8, respectively. The ratio is determined by the 20/80 principle~\cite{SYY20} as suggested in the previous study~\cite{HZC22} to avoid unreliable relation constructions. Following previous studies~\cite{QWL21, JGP16,ZDZ22}, we search the initial temperature parameter $\tau$ in $\{10^{-2},10^{-1},1,10,10^2,10^3\}$ and anneal it after every 40 batches. The hyper-parameters of all competing models either follow the suggestions from the original papers or are carefully tuned, and the best performances are reported. We implement
\model in PyTorch 1.7.1, Python 3.8.3, and RecBole v1.0.1~\cite{ZMH21} on a workstation with an Intel Xeon Platinum 2.40GHz CPU, a NVIDIA Quadro RTX 8000 GPU, and 754GB RAM. The source code is available online at \url{https://github.com/zc-97/SSDRec}.

\subsection{Performance Comparison (RQ1 and RQ2)}

We present the main experimental results in Table~\ref{tab:Overall-comprehension-1} and Table~\ref{tab:Overall-comprehension-2}.
All improvements are significant by performing a two-sided $t$-test with $p<0.05$ over the strongest baselines. We can draw a few key observations as follows:

\begin{itemize}[\IEEElabelindent=0pt]
    \item When integrated into different sequential recommendation methods, the proposed \model framework can consistently achieve the best performance on all datasets as shown by the average improvements of all metrics. Its average relative improvements over SASRec are 110.64\%, 39.70\%, 39.34\%, 51.89\%, and 23.40\% on ML-100K, ML-1M, Beauty, Sports, and Yelp, respectively. After applying  \model for sequence denoising, the performance of mainstream sequential recommenders is significantly improved. These results generally demonstrate the flexible applications and effectiveness of our solution in boosting the performance of various mainstream sequential recommendation methods.
    \item Compared with applying \model to RNN-based and CNN-based sequential recommenders, applying \model to Transformer-based sequential models can achieve even larger performance boost. 
    Transformer-based methods (i.e., dot-product attention mechanism) have to learn an attention matrix based on all items in a sequence. Although the attention mechanism can assign lower attention scores to noisy items, the raw sequence's noise still inevitably affects the learned attention scores. Through the proposed \model model, these methods can eliminate noise before calculating attention scores and thus assign higher weights to noiseless items, leading to better representations.
    \item Compared with the state-of-the-art denoising methods and debiased method, \model consistently yields the best performance on all datasets. For example, its relative improvements over the strongest baselines are 22.91\%, 22.87\%, 9.51\% in terms of HR@K, 16.44\%, 22.65\%, 12.17\% in terms of NDCG@K, and 18.37\% in terms of MRR on ML-100K. Such results demonstrate the superiority of \model.  
    \item Compared with FMLP-Rec, an implicit denoising method, explicit denoising methods (e.g., DSAN) can achieve better performance in most cases, which aligns with our motivation that explicitly pinpointing noise and learning representations based on noiseless sub-sequences can boost sequential recommenders' performance.
    \item Compared with DCRec, a debiased sequential recommendation method, most sequence denoising methods cannot achieve better performance. The reason is that previous sequence denoising methods suffer from \problem, which inevitably leads to sub-optimal representation learning.
    \item Among the denoising methods and the debiased method, our proposed \model framework consistently achieves the best performance. We attribute such improvements to performing explicit sequence augmentation before sequence denoising to alleviate \problem, which can reliably identify noise in sequences and generate better sequence representations.  
\end{itemize}
In conclusion, the experimental results on five datasets demonstrate the flexible applications and superiority of the proposed \model model over various sequential recommendation models and state-of-the-art denoising methods.

\subsection{Ablation Study (RQ3)}
\label{sec:ablation study}
To verify the contribution of each stage of \model, we conduct an ablation study with various variants over the ML-100k dataset, including (1) \textit{w/o \model-1} using only the second and third stages, (2) \textit{w/o \model-2} (integrating HSD~\cite{ZDZ22} with \model-1) using only the first and third stages, and (3) \textit{w/o \model-3} using only the first and second stages.


Table~\ref{tab: ablation study} shows the performances of different variants in terms of HR@k, NDCG@K, and MRR@K. We can draw a few interesting observations as follows:
\begin{itemize}[\IEEElabelindent=0pt]
    \item It can be observed that each stage positively contributes to model performance. With the three-stage learning paradigm, \model can consistently outperform the other variants.
    \item The first stage of \model is the most crucial component. Excluding the first stage (i.e., encoding global relations) results in a significant performance drop. Without the learned global relations, we have no other prior knowledge (e.g., supervisory signals to reveal noise or additional features) to correctly guide subsequent item/position selections.
    \item Dropping the second stage of \model harms model performance. Compared with HSD~\cite{ZDZ22}, \model-2 can improve HSD's performance in all cases, demonstrating that learning global information can better sequence denoising performance. However, compared with \model-2, the proposed \model framework can significantly improve the performance. We attribute such improvements to the proposed self-augmentation module, which alleviates \problem.
    \item Excluding the third stage of \model could also harm the final performance. It confirms that \model can improve sequence denoising by removing false augmentations.
\end{itemize}
In conclusion, the above experimental results well justify 
our design choices. We observe similar trends on other datasets, which are omitted due to the space limit.

\begin{table}[t]
\caption{Impact of Different Stages of SSDRec on ML-100K dataset.}
\vspace{-1mm}
\label{tab: ablation study}
\resizebox{\linewidth}{!}{\begin{tabular}{lcccccc}
\toprule
Metrics      & HR@10          & HR@20          & N@10         & N@20         & MRR@10          & MRR@20          \\ \midrule
w/o SSDRec-1 & 0.0332          & 0.0707          & 0.0160          & 0.0255          & 0.0113          & 0.0141          \\
w/o SSDRec-2 & 0.0599          & 0.1178          & 0.0319          & 0.0427          & 0.0211          & 0.0241          \\
w/o SSDRec-3 & 0.0423          & 0.0885          & 0.0197          & 0.0294          & 0.0136          & 0.0162          \\ \midrule
HSD & 0.0583         & 0.1018          & 0.0225          & 0.0401          & 0.0202          & 0.0234          \\
SSDRec       &\textbf{0.0795} 	&\textbf{0.1336}	&\textbf{0.0379}	&\textbf{0.0507} 	&\textbf{0.0255} 	&\textbf{0.0290}  \\ \bottomrule
\end{tabular}}
\end{table}

\begin{table}[t]
\caption{Model Efficiency Study in Terms of Per-epoch Training and Inference Time (Seconds) on ML-100K, ML-1M, Beauty, Sports, and Yelp Datasets.}
\centering
\resizebox{\columnwidth}{!}{%
\begin{tabular}{clrrrr}
\toprule
Mode & \multicolumn{1}{c}{Dataset} & \multicolumn{1}{c}{HSD} & \multicolumn{1}{c}{STEAM} & \multicolumn{1}{c}{DCRec} & \multicolumn{1}{c}{SSDRec} \\ \midrule
\multirow{5}{*}{Training}     & ML-100K & 4.10   & 0.29   & 1.93  & 5.98   \\
                           & ML-1M   & 64.61  & 52.66  & 3.96  & 119.82 \\
                           & Beauty  & 67.19  & 47.42  & 8.77  & 126.62 \\
                           & Sports  & 148.36 & 113.70 & 13.75 & 245.08 \\
                           & Yelp    & 99.88  & 107.28 & 20.35 & 199.90 \\ \midrule
\multirow{5}{*}{Inference} & ML-100K & 3.02   & 0.02   & 3.83  & 0.88   \\
                           & ML-1M   & 7.74   & 96.90  & 0.41  & 10.34  \\
                           & Beauty  & 6.81   & 10.99  & 0.43  & 8.43   \\
                           & Sports  & 8.95   & 18.91  & 0.82  & 14.02  \\
                           & Yelp    & 9.90   & 19.93  & 0.86  & 13.34  \\ \bottomrule
\end{tabular}%
}
\label{tab: efficiency study}
\end{table}

\begin{figure}[t]
\centering
  \includegraphics[width=\linewidth]{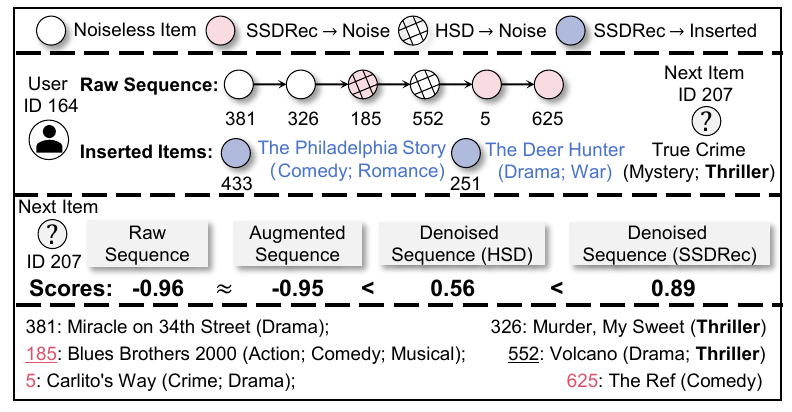}
 \vspace{-3mm}
\caption{A case study to show how the three-stage learning paradigm of \model affects next-item recommendation on ML-100K.}
\label{fig:case-study}
\end{figure}

\begin{figure*}[t]
\centering
  \includegraphics[width=\linewidth]{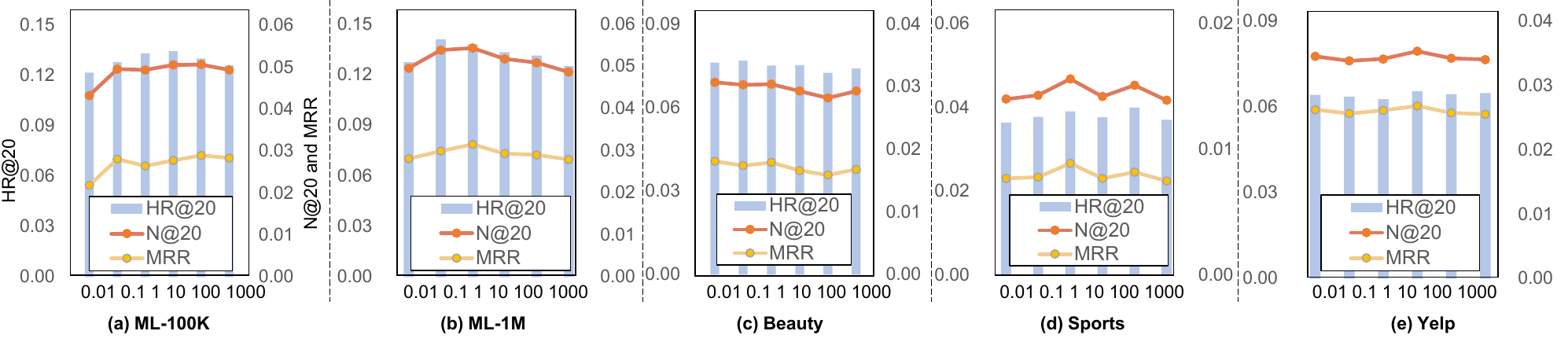}
 \setlength{\abovecaptionskip}{-2mm}
\caption{Hyperparameter study for \model in terms of HR@20, N@20, and MRR on ML-100K, ML-1M, Beauty, Sports, and Yelp datasets.}
\label{fig: hyper-parameter}
\end{figure*}

\subsection{Model Efficiency Study (RQ4)}
We investigate the efficiency of our \model framework as compared to the state-of-the-art denoising and debiased methods. We study the efficiency of all competing methods by evaluating their running time (seconds) of each training epoch and inference epoch on ML-100K, ML-1M, Beauty, Sports, and Yelp datasets. Table~\ref{tab: efficiency study} presents the evaluation results. 
\begin{itemize}[\IEEElabelindent=0pt]
    \item As can be seen, \model's computational costs are comparable to the explicit sequence denoising methods (e.g., HSD~\cite{ZDZ22} and STEAM~\cite{LWC23}), which aligns with our theoretical time complexity analysis.
    \item Recall that we implement \model by employing HSD in the third stage to generate denoised sequences. Thus \model's computational cost is higher than that of HSD. However, the extra computational cost of \model is acceptable. It is worth noting that the extra cost does not impact online inference. 
    This is because the designed self-augmentation module aims to enrich raw sequences. After training, the downstream denoising method can learn sufficient information. Therefore, we do not need to insert items during online inference.
    \item  Moreover, due to the possibility of encoding different relations in the global relation encoder independently, advanced parallel computing techniques can be employed to accelerate model training in practical recommendation scenarios. 
\end{itemize}

\subsection{Case Studies of \model's Explainability (RQ5)}
\label{subsec:case_study}
We conduct a case study to illustrate how the three-stage learning paradigm of \model can affect noise identifications and sequential recommendation. In Figure~\ref{fig:case-study}, we show a user whose ID is 164 and whose next-item ID is 207. The user has interacted with 42 items on the ML-100k dataset. For clarity and simplicity, we detail his recent six interactions, i.e., items $\{381,326,185,552,5,625\}$. 

\begin{itemize}[\IEEElabelindent=0pt]

    \item Learning sequence representations based on the raw sequence data yields the lowest score (i.e., $-0.96$) of recommending item 207 to the user. In contrast, after the second stage of \model, we select two items (i.e., items with IDs 433 and 251 denoted by blue circles) as the inserted items to perform sequence augmentation for the subsequent denoising process, yielding a similar score (i.e., $-0.95 \approx -0.96$) of recommending item 207, which aligns with our purpose that the proposed self-augmentation module can help better reveal noise without significantly changing the user preference embedded in the raw sequence.
    \item  After that, we identify three items denoted by red circles as noise and explicitly remove them from the raw sequence, thus generating the highest score (i.e., $0.89$) of recommending item 207 to the user. Moreover, compared with the denoised sequence of HSD~\cite{ZDZ22} (i.e., identifying item 185 and item 552 as noise denoted by circles with the `$\#$' symbol and yielding the score $-0.96<0.56<0.89$ of recommending the item with ID 207), \model additionally removes two items with IDs 5 and 625 (two comedies irrelevant to item 207) and keeps item 552 (a thriller relevant to item 207). In particular, for item 5 and item 625, HSD cannot identify them as noisy items. In contrast, through the devised self-augmentation model in the second stage of the proposed \model framework, our solution can explicitly pinpoint the noisy items and avoid under-denoising results. Moreover, \model can additionally keep some items (e.g., item 552) to prevent sequence denoising from the over-denoising problem, achieving better performance than HSD.
\end{itemize}
In conclusion, this case study reinforces our motivation that sequence denoising can help to learn better sequence representations. However, the limited information available in a single sequence could lead to \problem, which harms recommendation performance.
Moreover, after denoising, the interaction ratios we drop on ML-100K, ML-1M, Beauty, Sports, and Yelp are 24.22\%, 25.10\%, 26.28\%, 22.96\%, and 39.41\%, respectively, which shows \model's capability of eliminating noise.

\subsection{Hyperparameter Investigation (RQ6)}
Finally, to investigate the effect of different parameter settings, we
perform experiments to evaluate the performance of our proposed \model framework with different configurations of the important hyperparameter (i.e., $\tau$, used in Section~\ref{subsee:self-augmentaion}). When varying $\tau$, we keep other parameters with default values. The evaluation results in terms of HR@20, N@20, and MRR are presented in Figure~\ref{fig: hyper-parameter}. The x-axis represents different $\tau$ values. We summarize the results with the following observation: while the best $\tau$ values vary from dataset to dataset, the best $\tau$ values in small datasets (e.g., ML-100K) are typically lower than the values in large datasets (e.g., ML-1M and Yelp). Adopting a low $\tau$ value for initialization may exaggerate the denoising results in the early stage of the training process, resulting in worse performance.

\section{Related Work}
\label{sec:related_work}
Most sequential recommendation models~\cite{HKB15, DLZ17, TW18, LRC17, LZM18, KM18, SLW19, WTZ19, HZC22, DYG23, LSZ23} assume that users' interaction sequences are clean, which often does not hold in practice. When dealing with noisy sequences, their representation learning capabilities may be largely impacted. As a result, sequence denoising methods~\cite{YSS21, TWL21, SWS21, ZDZ22, ZYZ22,LWC23} have gained increasing attention in recent years. These methods mainly focus on attenuating the negative influence of noise from different perspectives. 


A key challenge of sequence denoising is the absence of supervised labels indicating noisy items. 
For this reason, existing sequence denoising approaches, typically operating under the self-supervised learning paradigm, normally assume that most interactions are noiseless and rely on detectable signals to identify and remove noise from raw sequences. We categorize the existing approaches into two major lines: implicit denoising (e.g., keeping noisy items but assigning them lower weights~\cite{ZYZ22}) and explicit denoising methods (e.g., completely removing noise~\cite{HZC22}). The first line implicitly mitigates noise's influence by maintaining noisy items in the sequence but assigning them lower importance in order to improve sequence representation learning. Despite its effectiveness, this approach may only partially mitigate the negative impact of noise and can still lead to inaccurate learning of user preferences, as evidenced by FMLP-Rec~\cite{ZYZ22}, which conducts denoising at the representation level rather than the item level.

The second line seeks to learn detectable signals (e.g., smooth sequentiality and correlations) in a sequence to identify noise. Unlike the first line, explicit denoising methods typically perform better since they remove most noise from sequences and provide noiseless sub-sequences for high-quality representation learning. The general intuition is that most items in a sequence exhibit these signals and, therefore, can be considered noiseless, while noisy items do not. However, designing and detecting these signals in sequences pose a critical challenge.
While these two types of denoising methods outperform traditional sequential recommendation approaches due to their denoising considerations, the inherent limitations of available information in a single sequence, especially in short sequences in many applications, may still result in unreliable noise identification. The learned sequential information may be insufficient to accurately denoise. Overlooking this limitation may result in \textit{over-denoising} and \textit{under-denoising} problems (\problem), which deserves an in-depth exploration.

Motivated by the observation, \model improves sequence denoising for sequential recommendation by executing explicit sequence augmentation prior to the denoising process, thereby alleviating \problem. It employs a three-stage learning paradigm that does not need additional features (such as extra user/item features beyond IDs or knowledge graphs) or noise's labels for sequential recommendation. This flexibility allows \model to be seamlessly integrated into existing denoising models and/or sequential models to enhance their learning capabilities.

Finally, it is worth noting that there are some recent denoising methods for other recommendation tasks, e.g., collaborative filtering~\cite{WFH21,GDH22}, multi-behavior recommendation~\cite{ZCZH23}, next-basket recommendation~\cite{QWL21}, and CTR prediction~\cite{LZW23}. These methods, unfortunately, cannot be directly applied to the sequence denoising problem for sequential recommendation.

\section{Conclusion}
\label{sec:conclusion}
In this paper, we studied the problem of sequence denoising for sequential recommendation from a new perspective -- how to solve the common over-denoising and under-denoising problems due to limited available information in a single sequence. We proposed to tackle \problem by performing sequence augmentation by incorporating global information before denoising. To effectively overcome the challenges of sequence augmentation, we consequently devised a novel \model framework with a unique three-stage learning paradigm. This framework incorporates a global relation encoder in its first stage, harnessing multi-faceted inter-sequence relations to establish prior knowledge. Consequently, we propose a self-augmentation module and a hierarchical denoising module in the second stage and third stage, respectively. 
Extensive experimental results showed that the \model framework can consistently outperform state-of-the-art methods. 

\section*{Acknowledgments}
This work was supported by the Heilongjiang Key R\&D Program of China under Grant No. GA23A915 and the National Natural Science Foundation of China under Grant No. 62072136. Xiangyu Zhao was supported by APRC - CityU New Research Initiatives (No. 9610565, Start-up Grant for New Faculty of City University of Hong Kong), CityU - HKIDS Early Career Research Grant (No. 9360163), Hong Kong ITC Innovation and Technology Fund Midstream Research Programme for Universities Project (No. ITS/034/22MS), Hong Kong Environmental and Conservation Fund (No. 88/2022), and SIRG - CityU Strategic Interdisciplinary Research Grant (No. 7020046, No. 7020074), Huawei (Huawei Innovation Research Program), Ant Group (CCF-Ant Research Fund, Ant Group Research Fund), Tencent (CCF-Tencent Open Fund, Tencent Rhino-Bird Focused Research Program), CCF-BaiChuan-Ebtech Foundation Model Fund, and Kuaishou. Peng Tang was supported by the National Natural Science Foundation of China under Grant No. 62002203, Shandong Provincial Natural Science Foundation No. ZR2020QF045, and Young Scholars Program of Shandong University.

\clearpage
\bibliographystyle{IEEEtran}
\bibliography{7Reference}

\end{document}